# Bayesian Optimization of High-Entropy Alloy Compositions for Electrocatalytic Oxygen Reduction

Jack K. Pedersen,[a] Christian M. Clausen,[a] Olga A. Krysiak,[b] Bin Xiao,[c] Thomas A. A. Batchelor,[a] Tobias Löffler,[b,c,d] Vladislav A. Mints,[e] Lars Banko,[c] Matthias Arenz,[a,e] Alan Savan,[c] Wolfgang Schuhmann,[b] Alfred Ludwig,[c,d] and Jan Rossmeisl*[a]

[a] J. K. Pedersen, C. M. Clausen, Dr. T. A. A. Batchelor, Prof. M. Arenz, Prof. J. Rossmeisl
Center for High Entropy Alloy Catalysis (CHEAC), Department of Chemistry
University of Copenhagen
Universitetsparken 5, 2100 København Ø (Denmark)
E-mail: jan.rossmeisl@chem.ku.dk

[b] Dr. O. A. Krysiak, Dr. T. Löffler, Prof. W. Schuhmann
Center for Electrochemical Sciences (CES), Faculty of Chemistry and Biochemistry
Ruhr University Bochum
Universitätsstr. 150, 44780 Bochum (Germany)

[c] Dr. B. Xiao, Dr. T. Löffler, L. Banko, A. Savan, Prof. A. Ludwig
Chair for Materials Discovery and Interfaces, Institute for Materials, Faculty of Mechanical Engineering
Ruhr University Bochum
Universitätsstr. 150, 44780 Bochum (Germany)

[d] Dr. T. Löffler, Prof. A. Ludwig
ZGH
Ruhr University Bochum
Universitätsstr. 150, 44780 Bochum (Germany)

[e] V. A. Mints, Prof. M. Arenz
Center for High Entropy Alloy Catalysis (CHEAC), Department of Chemistry, Biochemistry and Pharmaceutical Sciences
University of Bern
Freiestrasse 3, 3012 Bern (Switzerland)

**Abstract:** Active, selective and stable catalysts are imperative for sustainable energy conversion, and engineering materials with such properties are highly desired. High-entropy alloys (HEAs) offer a vast compositional space for tuning such properties. Too vast, however, to traverse without the proper tools. Here, we report the use of Bayesian optimization on a model based on density functional theory (DFT) to predict the most active compositions for the electrochemical oxygen reduction reaction (ORR) with the least possible number of sampled compositions for the two HEAs Ag-Ir-Pd-Pt-Ru and Ir-Pd-Pt-Rh-Ru. The discovered optima are then scrutinized with DFT and subjected to experimental validation where optimal catalytic activities are verified for Ag-Pd, Ir-Pt, and Pd-Ru binary systems. This study offers insight into the number of experiments needed for exploring the vast compositional space of multimetallic alloys which has been determined to be on the order of 50 for ORR on these HEAs.

## Introduction

High-entropy alloys (HEAs; in form of single-phase compositionally complex solid solutions) offer a vast composition space for optimization of catalytic properties,[1,2,3,4,5] because of the many multi-element atomic surface sites made possible on such complex surfaces contributing to a near-continuum of adsorption energies of the reaction intermediates involved in the catalytic transformations.[1] This is advantageous since new catalysts are especially needed to facilitate chemical reactions for sustainable energy conversion in order to meet the increasing global energy demand and to combat climate change.[6] One example of a key reaction in the hydrogen cycle is the oxygen reduction reaction (ORR), where current catalysts are still far from ideal and cannot meet the demands for commercially viable industrial implementation on a global scale and new innovations are highly seeked for.[7,8]

Combinatorial exploration of vast alloy composition spaces has been actively used as a tool in experimental catalyst discovery for a variety of reactions and constituent elements,[9,10,11,12,13,14,15,16] and efficient sampling of catalyst materials has also progressed.[17] However, when the number of constituent elements increases, the number of possible compositions grows combinatorially large and individual subsequent testing cannot be accomplished within realistic time scales (see supporting information (SI), Figure S2). This calls for the need to sample the composition space more efficiently as made possible by guiding the search with the aid of a surrogate function. Bayesian optimization of a Gaussian process (GP) surrogate function is a feasible choice for intelligent sampling problems,[18] and Bayesian optimization has also been employed to optimize the catalytic activity for methanol oxidation of a ternary alloy.[19] However, knowing beforehand how many experiments would be needed in such a compositional search is crucial for determining if such a search is tractable in the first place.

We propose a way to estimate the number of experiments needed using a model that has been found to correctly predict experimental trends for electrocatalytic ORR across hundreds of different alloy compositions within the Ag-Ir-Pd-Pt-Ru system.[4] Because of that, we expect the model to reproduce the catalytic complexity similar to an equivalent experimental search, and therefore it is likely to be suited as a proxy that can substitute most of the necessary experiments by simulations. By sampling alloy compositions from the model, the sampling of a real experiment can thus be emulated, and the number of experiments needed for future composition space optimization can be estimated. Modeling the catalytic activity of highly diverse and complex surfaces is still in its infancy with only a few studies conducted,[1,3,4,20,21] but modeling of other aspects relevant for catalysis, such as surface stability under reaction conditions is also being investigated.[22,23]

Using the Ag-Ir-Pd-Pt-Ru and Ir-Pd-Pt-Rh-Ru HEAs as exemplary systems for a composition optimization, we use the kinetic model combined with Bayesian optimization to suggest alloy compositions for which high catalytic activities for the ORR are predicted, while sampling as few compositions as possible to estimate the minimum number of experiments needed, and subsequently subject these optima to experimental validation. Moreover, by sampling the whole space of alloy compositions with the model in, say, 5 atomic percent (at.%) intervals, it is possible to assure with reasonable certainty that all local and global optimal compositions have indeed been identified by the Bayesian optimization.

## Results and Discussion

We apply our previously published model[4] for predicting current densities at 0.82 V vs. the reversible hydrogen electrode (RHE) on the face-centered cubic (fcc) (111) surfaces of the disordered quinary alloy systems of Ir-Pd-Pt-Rh-Ru and Ag-Ir-Pd-Pt-Ru. For construction of the model, thousands of *OH and O* adsorption energies were calculated with DFT in order to enable *OH and O* adsorption energy predictions on any surface site of the alloy at any composition (for details see SI). Due to the linear scaling between *OH and *OOH adsorption energies, focusing on those two intermediates is sufficient to predict the catalytic activity.[24] The model effectively maps an alloy composition to a relative measure of a current density at a given potential using equations 1-3. By doing so it takes as input net adsorption energies of on-top *OH and hollow site O* obtained by considering an intersite neighbor blocking effect that ensures that no neighboring on-top and hollow sites can adsorb intermediates at the same time (for details see SI).

$$j = \frac{1}{N} \sum_{i}^{N_{ads}} j_i \tag{1}$$

$$\frac{1}{j_i} = \frac{1}{j_D} + \frac{1}{j_{k,i}} \tag{2}$$

$$j_{k,i} = -\exp\left( - \frac{|\Delta G_i - \Delta G_{opt}| - 0.86 eV + eU}{k_B T} \right) \tag{3}$$

Here $j$ is the per site current density (in arbitrary units only used for comparing catalytic activity between compositions), $N$ is the number of surface atoms in the simulated surface, $N_{ads}$ is the number of sites at which adsorption has happened (after considering the intersite neighbor blocking), $j_i$ is the current at surface site $i$ modeled using the Koutecký-Levich equation, $j_D$ is the diffusion limited current (set to -1) ensuring that the current at each site only increases sigmoidally at high overpotentials, $j_{k,i}$ is the kinetically limiting current for site $i$ modelled using an Arrhenius-like expression assuming a Sabatier volcano relationship with the adsorption energies, $\Delta G_i$ is the *OH or O* adsorption free energy, $\Delta G_{opt}$ is the optimal *OH or O* adsorption free energy (set to 0.1 eV[25] and 0.2 eV[26] larger than for Pt(111) for *OH and O* respectively as suggested by theory and experiment), $e$ is the elementary charge, $U_{RHE}$ is the applied potential vs. RHE, $k_B$ is the Boltzmann constant, and $T$ is the absolute temperature (set to 300 K).

A surrogate function mapping alloy composition to current density was constructed as a GP with a squared exponential kernel scaled by a constant,

$$k(\boldsymbol{x}_i, \boldsymbol{x}_j) = C^2 \exp\left( - \frac{(\boldsymbol{x}_j - \boldsymbol{x}_i)^T (\boldsymbol{x}_j - \boldsymbol{x}_i)}{2\ell^2} \right) \quad , \tag{4}$$

where $\boldsymbol{x}_i$ and $\boldsymbol{x}_j$ are molar fraction vectors, and $C$ and $\ell$ are the constant value and length scale hyperparameters respectively of the GP which were optimized with every update of the sampled data. The superscript *T* denotes taking the transpose of the vector.

Two random compositions were initially chosen to form the basis for the Bayesian optimization. The expected improvement acquisition function was then used to suggest the next composition to investigate, taking into consideration the values and the readily obtained uncertainties of the surrogate GP.[27] The kinetic model was then used to compute the catalytic activity of the selected composition via Equations 1-3, and the surrogate GP was updated with this new sample using Bayesian inference as implemented in scikit-learn[28] (see the SI for details on the implementation). The process was repeated by letting the updated acquisition function choose the next composition of interest, and the optimization was allowed to run for 150 iterations which was generally found to be enough to discover the most active locally optimal compositions.

The quinary alloy composition space is equivalent to the set of all points in a 4-simplex (the 4 dimensional version of a regular tetrahedron), so plotting the resulting surrogate functions directly is hindered by the dimensionality of the plot. Instead, the local maxima obtained from the resulting surrogate functions for each of the quinary alloys are listed in Table 1, and for illustration a projection of the surrogate function for the Ag-Ir-Pd-Pt-Ru HEA in form of a pseudo-ternary plot with only the concentrations of Ag and Pd explicitly shown is given in Figure 1a for various stages of the optimization. Figure 1b shows the modeled current densities that are sampled during a run of the optimization with some noticeable minima (i.e. compositions with high absolute values of modeled current densities) shown explicitly as well as the emergence of the local minima of the surrogate GP. It is observed in Figure 1b that after discovery of a local minimum in the current density, the acquisition function ensures that new parts of the composition space is explored as evidenced from the typically sharp increases in the curve that follows. Figure 1c and d show the evolution of the constant value and length scale GP hyperparameters in Equation 4 as more compositions are sampled. Important to notice is the length scale of the kernel which, although not directly transferable to the compositions, does give an indication of the frequency with which the current density is expected to change with the compositions. To be specific, the found length scale of about 0.4 indicate that the current density is expected to vary with rather low frequencies as also indicated by the contours in Figure 1a, and therefore only a few local optima are expected for this hypersurface.

Indeed, the most active discovered optimal compositions in Table 1 form three groups of alloys, namely the binaries $Ag_{18}Pd_{82}$, $Ir_{\sim 50}Pt_{\sim 50}$, and the ternary $Ir_{\sim 10}Pd_{\sim 60}Ru_{\sim 30}$, with the latter two compositions discovered independently by both the Ag-Ir-Pd-Pt-Ru and the Ir-Pd-Pt-Rh-Ru quinary alloy models, supporting the robustness of the presented methodology when the input for the model is changed.

**Table 1.** Locally optimal compositions and the number of compositions needed to identify them for the two quinary HEAs.

| HEA | Local optimum[a] | Predicted current density (arb. units)[b] | Idenfication success rate[c] | Number of samples for identification of local optimum[b,d] |
|---|---|---|---|---|
| Ag-Ir-Pd-Pt-Ru | $Ag_{18}Pd_{82}$ | -0.203(2) | 100% | 50(21) |
| | $Ir_{9}Pd_{64}Ru_{27}$ | -0.160(2) | 100% | 28(28) |
| | $Ir_{48}Pt_{52}$ | -0.147(2) | 100% | 25(10) |
| | $Ag_{78}Ru_{22}$ | -0.063(3) | 69% | 93(27) |
| | $Ir_{46}Ru_{54}$ | -0.003(0) | 2% | 73(20) |
| | $Ir_{10}Ru_{90}$ | -0.002(1) | 14% | 110(27) |
| | Ru | -0.001(1) | 14% | 48(33) |
| Ir-Pd-Pt-Rh-Ru | $Ir_{42}Pt_{58}$ | -0.165(2) | 100% | 23(8) |
| | $Ir_{12}Pd_{56}Rh_{4}Ru_{28}$ | -0.164(1) | 100% | 19(10) |
| | Rh | -0.001(2) | 27% | 48(42) |

[a] Determined as the local optima of the resulting surrogate GP after sampling of 150 compositions for 64 random realizations of the two initial compositions (one such realization is shown in Figure 1). The spread in these compositions is on the order of 1 at.%. [b] Given as the mean followed by the sample standard deviation on the last digit(s) in parentheses. [c] Determined as the proportion of the resulting surrogate GPs after sampling of 150 compositions for 64 random initializations that identify the optimum as a local maximum. [d] Determined as the number of samples needed for those of 64 surrogate GPs with random initializations that successfully identified the optimum. The optimum has been considered identified when the molar fraction is within a 10 at.% difference from the optimum, e.g. $Ag_{23}Pd_{77}$ would be regarded as a successful discovery of the $Ag_{18}Pd_{82}$ optimum.

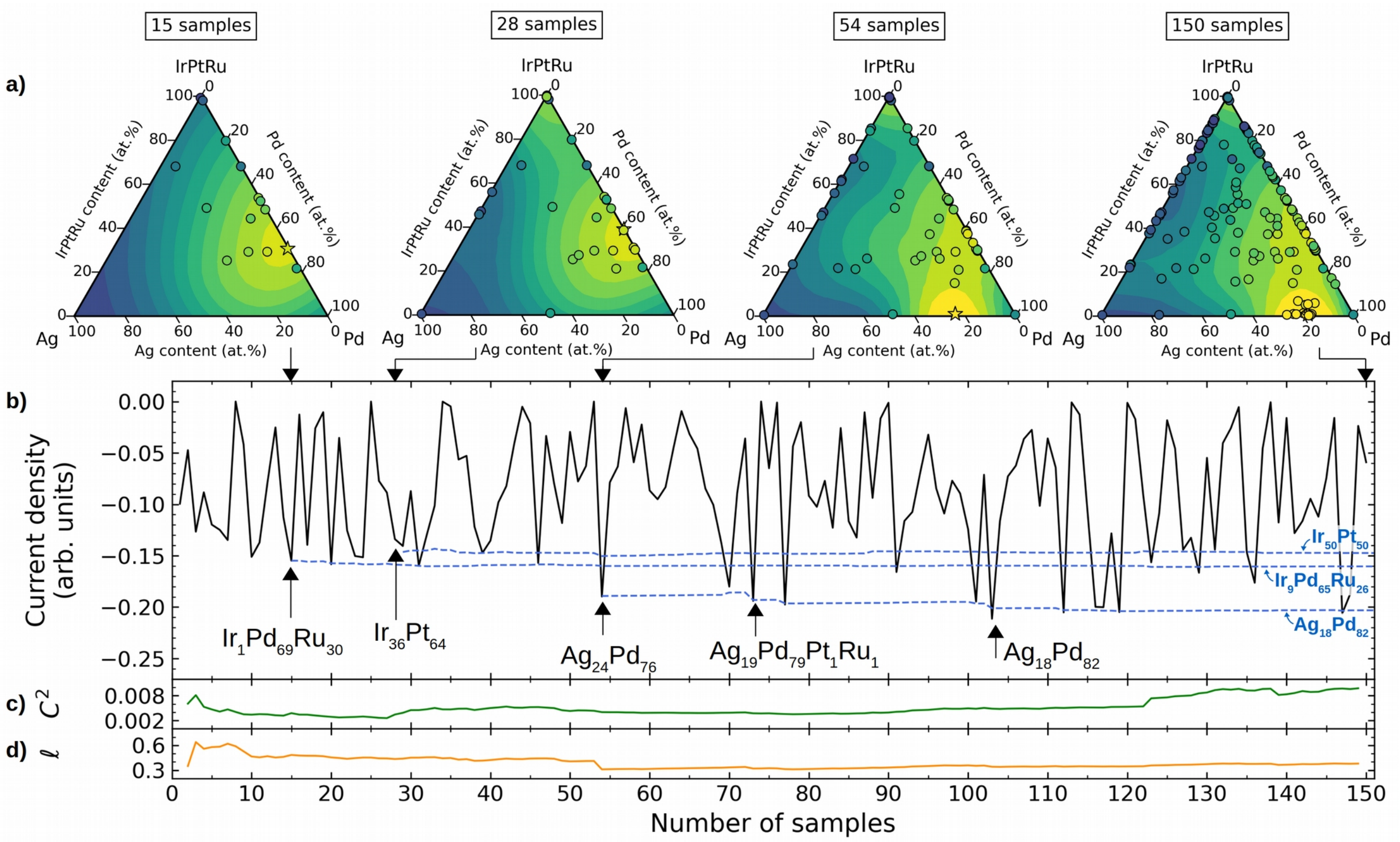

**Figure 1.** Example of a Bayesian ORR composition optimization for the quinary Ag-Ir-Pd-Pt-Ru system. **(a)** Pseudo-ternary plots (with Ir, Pt, and Ru collected into a single concentration) of the surrogate GP after sampling of 15, 28, 54, and 150 compositions. Yellow colors signify regions with high absolute values of modeled current density, and blue colors signify regions with correspondingly low values. Previously sampled compositions are shown as black circles, and the best composition found so far is marked with a star. When projecting current densities from the quinary to the pseudo-ternary composition space, more compositions will inevitably occupy the same points in the diagram. In the shown plots the maximal absolute value of the current density for overlapping compositions has therefore been depicted. **(b)** Current densities sampled during the Bayesian optimization (black solid line) and the emergence of the three most active locally optimal compositions (blue dashed lines). **(c,d)** Variation of the GP squared exponential kernels' (Equation 4) constant term (**c**) and length scale (**d**) hyper-parameters.

For proof of concept and to verify that all local optima had indeed been discovered by the Bayesian optimization, we simulated all compositions in 5 at.% for both HEA systems, corresponding to 10,626 simulations for each. This is a much more demanding task compared to Bayesian optimization, and without an automated setup it is also an impractical objective for an actual experimental realization, not to mention the cost associated with the precursor materials and the experimental setup.

The discovered locally optimal compositions using this 5 at.% grid search of the quinary composition space are shown in Table 2. It indeed appears that the most important compositions for catalysis were found by the Bayesian optimization. Most noticeably, the locally optimal compositions $Ag_{20}Pd_{80}$ and $Ir_{\sim 50}Pt_{\sim 50}$ with high expected current densities are confirmed, but it is also possible to match the other optima found by Bayesian optimization in Table 1 with corresponding counterparts in the 5 at.% grid search analysis in Table 2, e.g. the closely related $Ir_9Pd_{64}Ru_{27}$ optimum for the Ag-Ir-Pd-Pt-Ru HEA and the $Ir_{12}Pd_{56}Rh_4Ru_{28}$ optimum for the Ir-Pd-Pt-Rh-Ru HEA with a modeled current density of about -0.16 (arb. units) in Table 1 correspond to the $Ag_{\sim 0}Ir_{\sim 10}Pd_{\sim 60}Pt_{\sim 0}Ru_{\sim 30}$ and $Ir_{\sim 10}Pd_{\sim 60}Pt_{\sim 0}Rh_{\sim 0}Ru_{\sim 30}$ groups of compositions with similar current density highlighted in boldface in Table 2. In fact, the optima found with Bayesian optimization may fruitfully be thought of as the locally optimal composition that the compositions from the 5 at.% grid search would converge to if the 5 at.% step constraints were relaxed. We note that the trace amounts of other elements in the group of Pd-Ru-rich optima appear not to be very influential on the modeled current density as the quinary composition space forms a rather flat plateau around $Pd_{65}Ru_{35}$ as shown in Figure S4. We shall therefore simplify our analysis of this optimum in the following and treat it as a binary Pd-Ru alloy.

Similarly, the less optimal $Ag_{78}Ru_{22}$, can be assigned a counterpart ($Ag_{85}Ru_{15}$) in Table 2. However, it also appears that local optima of low absolute current densities are not matched as well between the Bayesian optimization and the 5 at.% grid search. For instance, the similar compositions $Ag_5Ir_{10}Pd_{20}Pt_{65}$ (from Ag-Ir-Pd-Pt-Ru) and $Ir_{15}Pd_{20}Pt_{55}Ru_{10}$ (from Ir-Pd-Pt-Rh-Ru) with a relatively high concentration of Pt at a modeled current density of around -0.09 (arb. units) from the 5 at.% grid search in Table 2 do not have counterparts from the Bayesian optimization in Table 1. The reason for this is either that these compositions are not actually local minima (if the 5 at.% steps were relaxed) and therefore not identified as such by the Bayesian optimization, or that they are indeed local minima but were not sampled by the acquisition function because they were predicted to have small absolute current densities. In either case this highlights the trade-off between exploratory sampling of the composition space and the exploitation of known catalytically active compositions which is characteristic for Bayesian optimization and will reduce sampling of suboptimal alloys.

**Table 2.** Locally optimal compositions found using a 5 at.% grid search over the two quinary composition spaces.

| HEA | Local optimum[a] | Modeled current density (arb. units) |
|---|---|---|
| Ag-Ir-Pd-Pt-Ru | $Ag_{20}Pd_{80}$ | -0.21 |
| | **$Ag_5Ir_{10}Pd_{60}Pt_5Ru_{20}$** | -0.16 |
| | **$Ag_5Ir_5Pd_{65}Ru_{25}$** | -0.16 |
| | **$Ir_{20}Pd_{60}Ru_{20}$** | -0.16 |
| | **$Pd_{65}Pt_5Ru_{30}$** | -0.16 |
| | **$Pd_{55}Pt_{20}Ru_{25}$** | -0.15 |
| | $Ir_{55}Pt_{45}$ | -0.15 |
| | $Ir_{45}Pt_{55}$ | -0.15 |
| | $Ag_5Ir_{10}Pd_{20}Pt_{65}$ | -0.09 |
| | $Ag_{85}Ru_{15}$ | -0.06 |
| Ir-Pd-Pt-Rh-Ru | **$Ir_{10}Pd_{55}Rh_5Ru_{30}$** | -0.17 |
| | $Ir_{50}Pt_{50}$ | -0.17 |
| | $Ir_{40}Pt_{60}$ | -0.17 |
| | **$Ir_{15}Pd_{60}Pt_5Ru_{20}$** | -0.16 |
| | $Ir_{15}Pd_{20}Pt_{55}Ru_{10}$ | -0.08 |
| | $Pt_{65}Rh_{35}$ | -0.07 |

[a] Defined as compositions for which a ±5 at.% change in any molar fraction would result in a less active catalyst. Compositions in **boldface** refer to the group of $Ir_{\sim 10}Pd_{\sim 60}Ru_{\sim 30}$ compositions with similar predicted current densities.

Since the model is initially fed calculated adsorption energies of the five-dimensional quinary alloys, it is essentially extrapolating to the two-dimensional *edges* of the composition space when predicting the activity of the discovered optimal compositions which appear to group mainly into the binary systems Ag-Pd, Ir-Pt, and Pd-Ru. This is because only the most central composition space is likely to be sampled when generating random configurations of the five elements in density functional theory simulated surfaces, even when sampling smaller unit cells like 2x2 atom-sized surfaces.

To confirm the local activity maxima and the model's predictive ability, DFT adsorption energies computed solely on the Ag-Pd, Ir-Pt and Pd-Ru binary alloys were used as input for the model and used to predict the current densities along those binary composition spans. Random atomic configurations from different molar ratios along these spans and adjusted sample weights assured a thorough representation of the two-dimensional composition space.

Additionally, several regression algorithms were tested against different truncations of the feature vectors (i.e. the extent of the considered adsorption site atomic environment). This confirmed that a combination of a per-unique-site-based linear regression model and the most influential neighboring atoms[29,30] maintains a low model complexity while still providing high adsorption energy prediction accuracy. However, for the binary alloys it was possible to achieve exceptionally low prediction error by using a non-linear regression algorithm (a gradient boosted decision tree) and an extensive description of the adsorption site motif, even though the 2x2 atom-sized surface holds limited information due to the periodic boundary conditions (see SI for details).

Plotting the current density output of the binary trained gradient boosted model against the same results from the linear model trained on the quinary alloys, it is seen for the Ag-Pd system in Figure 2 that at high Pd content both models predict an activity maximum around $Ag_{15}Pd_{85}$. This activity stems almost exclusively from O* bound in fcc hollow sites composed of two Pd atoms and one Ag atom with some contribution from three-fold Pd sites as shown in Figure 2b and c. However, due to the discrepancy of the models' prediction of *OH bound at on-top Pd sites (Figure 2d and 2e) the binary trained model retains high activity for a wider span of compositions compared to the quinary trained model which drops below the activity of Pt(111) at around 45 at.% Ag content as seen in Figure 2a, and thus we would still expect an appreciable catalytic activity for Ag-Pd at the equimolar ratio using the present model.

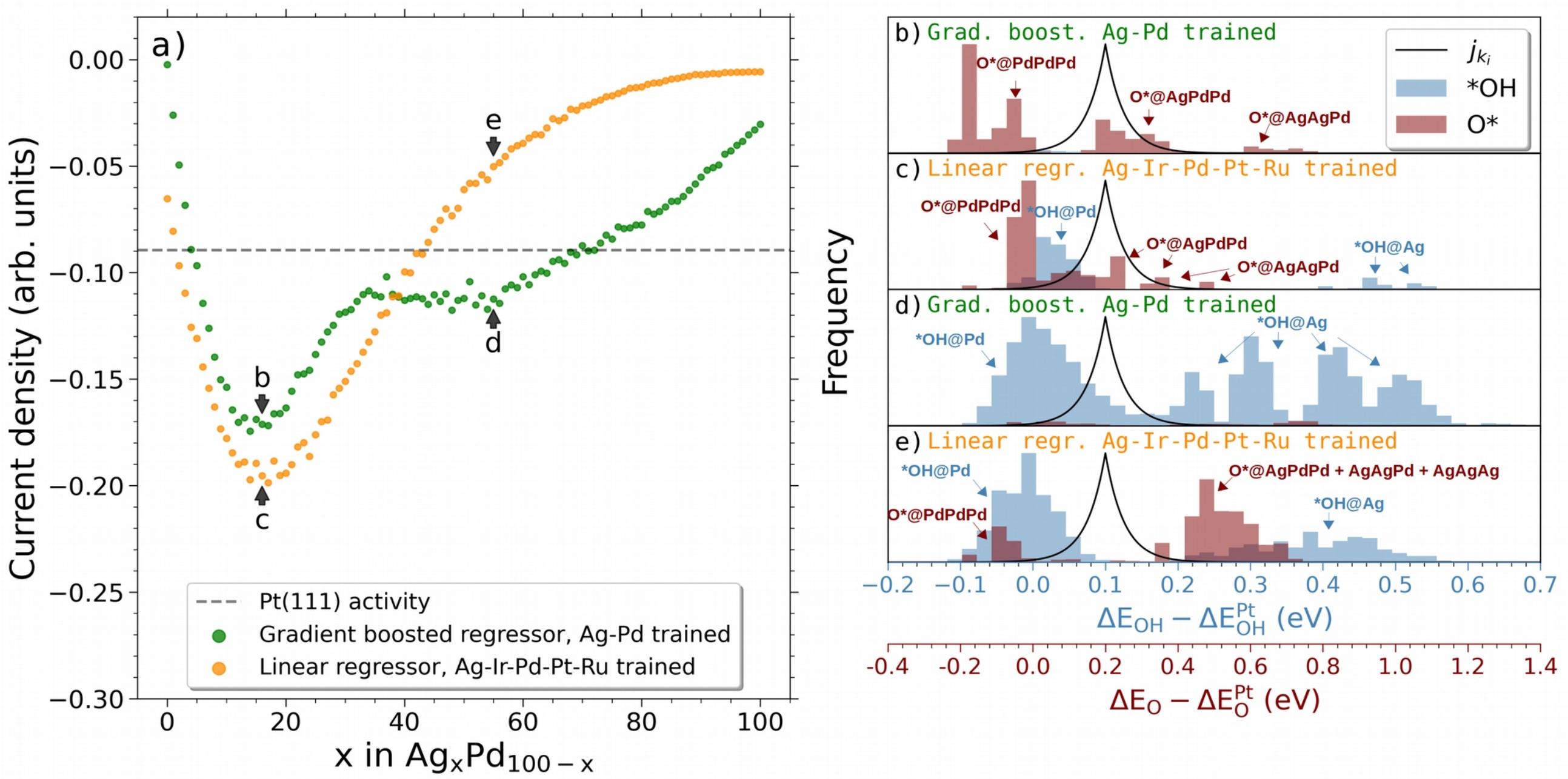

**Figure 2. a)** Simulated current densities of the Ag-Pd system in a composition range from pure Pd to pure Ag with 1 at.% increments. A linear regression model trained on DFT-calculated samples of the Ag-Ir-Pd-Pt-Ru alloy is used alongside a gradient boosted model trained on DFT calculated samples of Ag-Pd to predict the adsorption energies of the simulated surface. These predictions serve as input for Equations 1-3 which yield the resulting current densities. **b-e)** Net adsorption energy distributions (after intersite blocking) are displayed for selected compositions. The histograms are normalized against the largest bincount and a scaled visualization of the modeled current density in Equation 3 is shown.

Ir-Pt and Pt-Ru were similarly scanned as shown in Figure 3 and more detailed in Figures S7 and S8 and display overall good agreement with the quinary model. Both models predict optimum compositions around $Ir_{45}Pt_{65}$ and $Pd_{65}Ru_{35}$ and equivalent trends in the predicted current densities for the entire composition spans are observed.

The high catalytic activity of the Ag-Pd, Ir-Pt, and Pd-Ru alloys is not surprising since alloying the active elements Pd and Pt is a general way to enhance the activity for ORR[31] and since these alloys have indeed been tested experimentally with optimal compositions determined for the respective reaction conditions to be around $Ag_{10}Pd_{90}$,[32] $Ir_{15}Pt_{85}$,[33], and $Pd_{50}Ru_{50}$.[34]

To validate and compare the proposed catalytic trends of the discovered binary compositions, the ORR activity of thin-film composition spreads of the predicted Ag-Pd, Ir-Pt, and Pd-Ru binary alloys were synthesized and then analyzed by the use of a scanning droplet cell (SDC) in 0.1 M $HClO_4$. This high-throughput electrochemical technique allows localized characterization of selected compositions along the compositional gradient. Precise positioning above the investigated sample is enabled by assembling the SDC head with robotic arms and a force sensor. An electrochemical cell is formed by pressing the Teflon tip to the surface of the sample, defining the surface of the working electrode in every measurement area (MA). We used an automated setup to exclude any human error and provide the same measuring conditions for each measurement, together with the high reliability and accuracy of the data, which allows credible comparison of the activity between different MAs and samples. Figure 3 shows measured current density values vs. the composition of the Ag-Pd, Pd-Ru, and Ir-Pt systems. All linear sweep voltammograms (LSVs) are available in Figure S9. For the Pd-Ru composition spread we observed a broad minimum in ORR current densities for compositions ranging from ca. $Pd_{68}Ru_{32}$ to $Pd_{59}Ru_{41}$, covering the predicted optimal composition of $Pd_{65}Ru_{35}$. In case of the Ag-Pd composition spread, only a part of the compositions, those with lower Ag content, could be measured without visible corrosion (see Figure S9a). Here, a current maximum was found at the composition $Ag_{14}Pd_{86}$, corresponding very well to the predicted composition of $Ag_{15}Pd_{85}$.

On the contrary, the Ir-Pt composition spread shows a clear increase of the activity toward ORR with decreasing content of Ir, i.e. here we do not observe an agreement with the predicted optimal composition. In order to examine if the plateau at low Ir content observed in Figure 3c is the expected optimum, a sample covering higher Pt contents was prepared and tested as shown in Figure 3d.

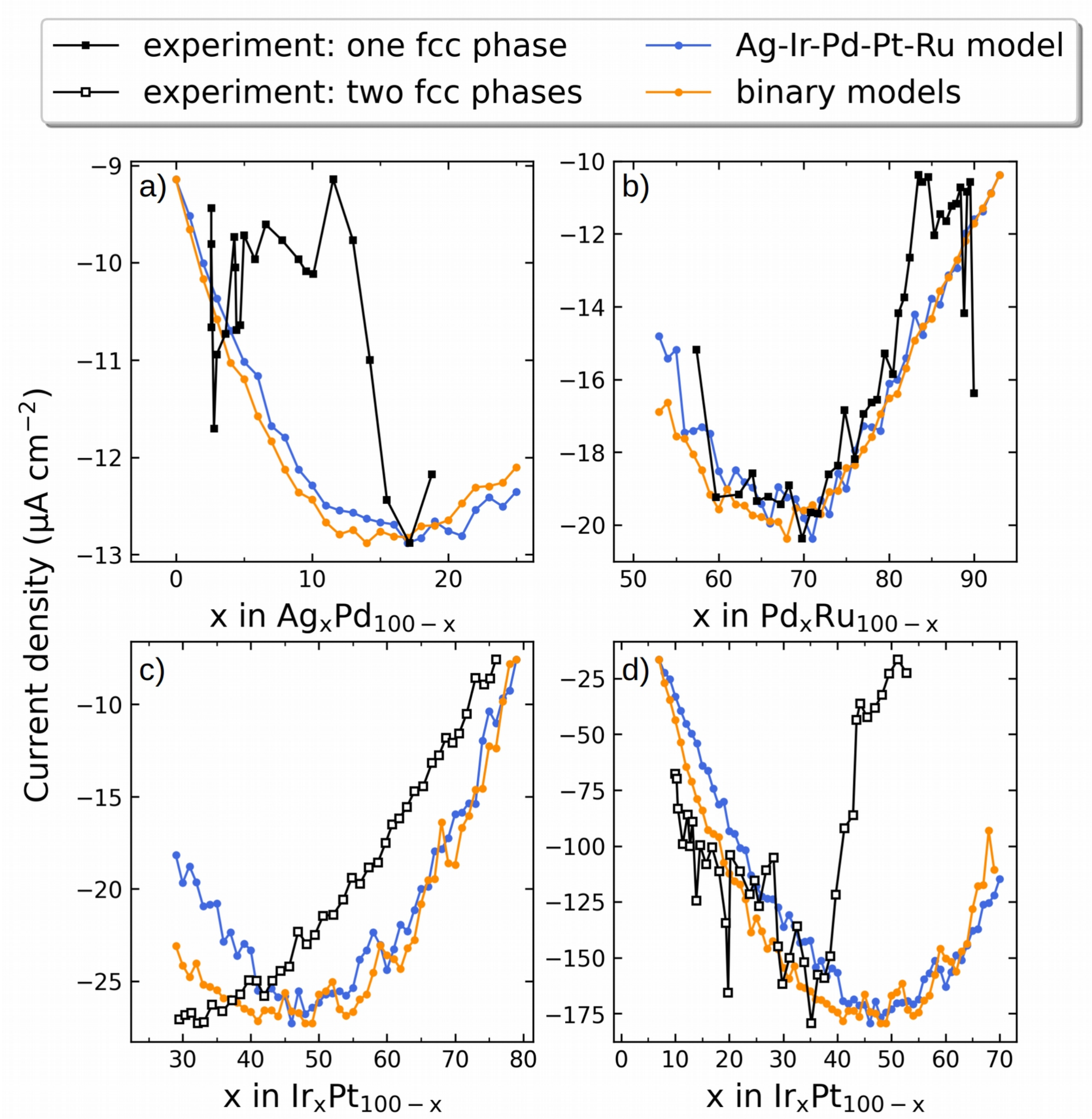

**Figure 3.** Comparison of simulated and experimental catalytic activities (black curves) for **a)** Ag-Pd, **b)** Pd-Ru, and **c,d)** Ir-Pt (for different composition ranges in **c)** and **d)**) at 800 mV vs. RHE. The simulated current densities were normalized to the experimental current densities by ensuring that the minimum and maximum current densities match up. For Pd-Ru in **b)** three outliers that gave rize to very high current densities were left out (see Figure S9b).

To ensure the enhanced activity can indeed be fully assigned to the composition effect and the impact of changes in surface roughness can be ruled out, atomic force microscopy (AFM) measurements of surface roughness in different spots of the binary thin-film composition spreads were used. For all of the considered samples the changes in surface roughness between different areas of the sample are negligible. Correlation of measured current densities with composition of the samples and their surface roughness are presented in Figure S10.

In order to determine the crystal structure and fully understand the measured correlation between current densities and binary compositions, X-ray diffraction (XRD) was conducted. The crystal structures of as-deposited Ag-Pd, Pd-Ru and Ir-Pt thin films for various compositions were determined from XRD diffractograms shown in Figure S11. Five XRD patterns were observed in all three binary systems, which are characteristic for Bragg reflections from fcc structures. The diffraction patterns exhibit the highest intensity reflection along the (111) plane and four weak reflections along the (200), (220), (311) and (222) planes. For Ag-Pd and Pd-Ru, the diffraction peaks continuously shift to lower 2θ values with increasing Ag or Pd amount. The lattice parameters for these two systems are determined from Bragg's law, and the calculated results show that the variation of lattice parameters with chemical composition agrees well with a linear dependence based on Vegard's law. This reveals that as-deposited Ag-Pd and Pd-Ru binary systems with different compositions form continuous solid-solutions with an fcc structure. In the case of the Ir-Pt system, the (111) peak splits into two peaks. This is due to the coexistence of two phases and implies that Ir and Pt could not be completely mixed, which is consistent with the large miscibility gap of the Ir-Pt phase diagram.

## Conclusion

To summarize, we have combined a kinetic model with Bayesian optimization to predict compositions of highest current density for ORR starting from the quinary HEAs Ag-Ir-Pd-Pt-Ru and Ir-Pd-Pt-Rh-Ru. The most important locally optimal compositions come out at around $Ag_{15}Pd_{85}$, $Ir_{50}Pt_{50}$, and $Ir_{10}Pd_{60}Ru_{30}$. The model, trained on DFT calculated *OH and O* adsorption energies on sites of the equimolar Ag-Ir-Pd-Pt-Ru and Ir-Pd-Pt-Rh-Ru HEAs, was successful in extrapolating catalytic activity trends to the discovered optimal binary systems as confirmed by training on new data specific for these alloys. The model was also shown to reasonably reproduce the catalytic activity trend from synthesized thin-film compositions spreads of the Ag-Pd, Ir-Pt, and Pd-Ru systems for which optimal compositions of around $Ag_{14}Pd_{86}$, $Ir_{35}Pt_{65}$, and $Pd_{65}Ru_{35}$ were determined. A direct comparison between the model and the experiment, however, should be done with caution since many reaction condition parameters are not accounted for in the model. While suggesting optimal alloy catalysts, the model is at the same time able to estimate the number of experiments needed for the discovery of optimal compositions in the vast compositional space of quinary alloy systems. With the Bayesian optimization of the kinetic model employed herein the number of experiments comes out at about 50 for discovery of the most important optima for the two investigated quinary HEAs. This number gives hope that experimental composition optimizations of vast multi-metallic composition spaces is indeed realizable.

## Acknowledgements

J.P., C.C., V.M., T.B., M.A., and J.R. acknowledge support from the Danish National Research Foundation Center for High-Entropy Alloy Catalysis (CHEAC) DNRF-149. J.P. acknowledges support from the Danish Ministry of Higher Education and Science (Structure of Materials in Real Time (SMART) grant), T.B. acknowledges support from VILLUM FONDEN (research grant 9455), W.S. acknowledges funding from Deutsche Forschungsgemeinschaft (DFG) under Germany's Excellence Strategy (EXC 2033 – 390677874 – RESOLV) and from the European Research Council (ERC) under the European Union's Horizon 2020 research and innovation programme (grant agreement CasCat [833408]. A.L. and B.X. acknowledge funding from DFG project LU1175/26-1. ZGH at RUB is acknowledged for using its experimental facilities.

## Conflict of interest

The authors declare no conflict of interest.

# Supporting Information

## Computational Procedures

All simulated structures, data and scripts necessary for reproducing the simulations herein have made freely accessible at https://nano.ku.dk/english/research/theoretical-electrocatalysis/katladb/bayesian-optimization-of-hea/

**Density functional theory simulations**

Density functional theory using the revised Perdew-Burke-Ernzerhof (RPBE) exchange-correlation functional[1] as implemented in the GPAW code[2,3] was used to obtain *OH and O* adsorption energies on fcc (111) 2x2-atoms-sized, four-layered surface slabs that were periodically repeated in the direction parallel to the slab. The structures were set up and manipulated in the Atomic Simulation Environment (ASE)[4]. The slabs were constructed with an fcc lattice constant set to the weighted average of the calculated fcc lattice constants of the elements in the surface layer, a vacuum of 7.5 Å was added above and below the slab, and the atoms in the two bottom layers were held fixed during geometry relaxations at which the structures were optimized until the maximum force on any atom was below at least 0.1 eVÅ$^{-1}$. The wave functions were expanded in plane waves with an energy cutoff set to 400 eV, and the Brillouin zone was sampled with a Monkhorst-Pack grid of 4x4x1 k-points. For training the Ag-Ir-Pd-Pt-Ru quinary alloy adsorption energy regressor, a total of 1304 *OH and 1768 O* adsorption energies were simulated on slabs where the metals in the structure were randomly sampled from an equimolar ratio. Equivalently for the Ir-Pd-Pt-Rh-Ru system where 856 *OH and 997 O* adsorption energies were simulated. The 2x2-atoms-sized, four-layered slabs for the binary alloy systems (Ag-Pd, Ir-Pt and Pd-Ru) were calculated with a similar computational setup with the exception of 10 Å added vacuum. In addition to slabs sampled from equimolar ratios, 25% of the slabs were sampled from a 3:1 ratio e.g. $Ag_{75}Pd_{25}$ and 25% from a 1:3 ratio e.g. $Ag_{25}Pd_{75}$ in order to more sensibly span the compositions of the binary alloy systems.

**Adsorption energy prediction**

The DFT calculated *OH and O* adsorption energies were used to train a regressor for predicting the *OH and O* adsorption energy at any conceivable on-top and fcc hollow site of an fcc (111) surface. To this end, we applied our previously developed scheme[5] for mapping structures into machine readable features simply by one-hot encoding the identity of the adsorption site ensemble and by counting the number of each element in equidistant positions from the adsorption site. For *OH on-top adsorption the atoms included in our description was the on-top adsorbing atom itself, the surface and subsurface neighboring atoms, and the second-nearest atoms in the third layer as we suggested recently[6]. For O* fcc hollow site adsorption the three-atom site ensemble as well as its surface and subsurface neighboring atoms were included in the description of the site (for an example see Figure S3). These features were used to fit a linear model for each on-top *OH adsorption site, i.e. one for on-top Ag, one for on-top Ir, etc. containing 15 fitted parameters each, as well as a single linear model for fcc hollow adsorbing O*, containing 55 fitted parameters (the fitted linear parameters are shown in Table S2-S5). For the investigated binary alloys a gradient boosted regressor was used on a more elaborate description of the surface site including the neighboring atomic environment up to the third or fourth closest atoms of each layer in order to improve on the prediction accuracy. Since the 2x2-atoms-sized DFT simulated slabs are periodically repeated this will include some zones without any additional information, however all available atoms will be included in the site features. In addition, a gradient boosted model was fitted to each O* adsorption site, i.e. one model for $Ag_3$, $Ag_2Pd$, $AgPd_2$ and $Pd_3$, respectively. Furthermore, when training the gradient boosted regressor the samples of binary alloys were weighted to enhance their representation of the composition span. Thus, each data set of a binary alloy contained two pure metal samples with assigned weights of 103, approximately 1000 samples drawn equally from the 1:3 and 3:1 molar ratios with assigned weights of 2 and approximately 1000 samples drawn from the equimolar ratio with assigned weights of 1. Linear and gradient boosted regression algorithms were used as implemented in scikit-learn[7] with default hyperparameters.

**Current density modeling**

The current density predicted by simulations was predicted using equations 1-3. However in order to improve on the model's predictive trends a simple adsorbate interaction between adsorbed *OH and O* was included.[8] This interaction works by ensuring that no two neighboring on-top and hollow sites can adsorb reaction intermediates at the same time. To calculate the current density using equation 1-3, the net coverage and corresponding net adsorption energies were used by accounting for this interaction. In practice the net adsorption energies were achieved by predicting the *OH and O* adsorption energies on all on-top and fcc hollow sites on a randomly constructed surface with a desired composition and measuring 100x100 atoms (the dependence of the predicted current density on the surface size is shown in Figure S1). The surface was then filled with *OH on-top and O* fcc hollow adsorbates starting at the strongest adsorbing sites and filled using the rule that no neighboring on-top and hollow sites can adsorb at the same time, until no more free surface sites remained. The net coverage and net adsorption energies of *OH and O* achieved in this way would then act as the input for equation 1-3 when calculating the predicted current density for the given composition.

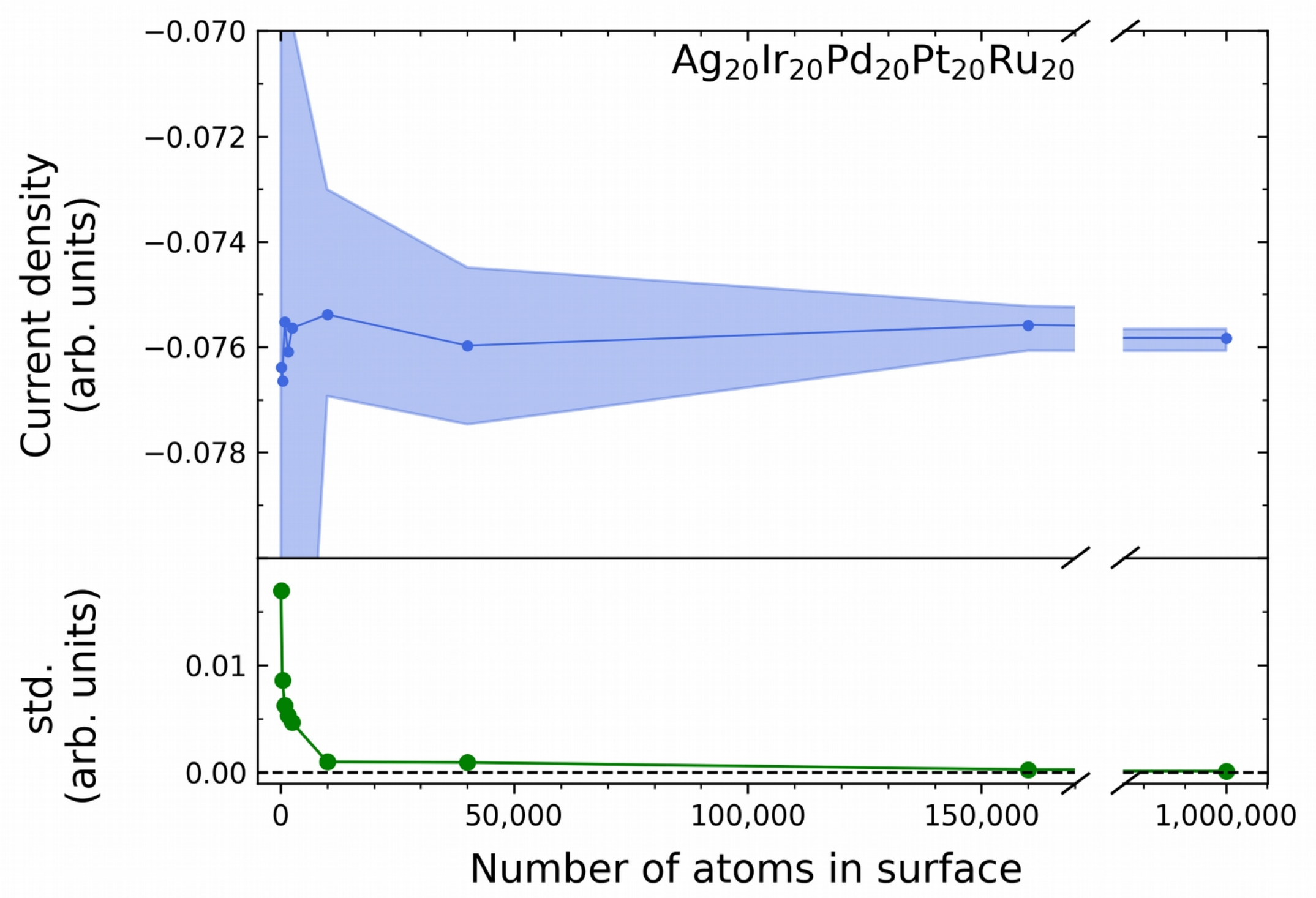

**Figure S1.** The predicted current density vs. the number of atoms of the simulated surface sampled for at least five random surfaces of equimolar AgIrPdPtRu at each of the sampled sizes. The standard deviation (std.) of the five sampled points are shown in green in the lower plot. At 10,000 atoms in the surface (100x100 atoms) the variation in the simulated current density is appreciably low compared to the variation between compositions (see for instance figure 2 in the main text).

**Bayesian Optimization**

In order to find optimal compositions with Bayesian inference, a Gaussian process regressor as implemented in scikit-learn[7] was initially trained on two randomly selected molar fractions along with their corresponding simulated current densities obtained with the kinetic model. The Gaussian process regressor was then used to predict current densities and surrogate model uncertainties at 1000 randomly selected molar fractions in order to span the quinary composition space in an approximate manner. The selection of the most optimal composition to sample next was performed with the *expected improvement* acquisition function. The principle of this acquisition function is to evaluate the expectation value of the improvement function,[9]

$$\mathrm{E}[I(\boldsymbol{x})]=\mathrm{E}[\max(y_{min}+\xi-Y,0)] \qquad \text{(S1)}$$

at a molar fraction $\boldsymbol{x}$ for randomly distributed current densities $Y$ and for the highest absolute value of the current density $y_{min}$ sampled by the kinetic model so far. $\xi$ is a tunable parameter that effectively adjusts $y_{min}$. If $\xi$ is chosen to be greater than zero, the minimum found so far is effectively increased making molar fractions with greater probability of having current densities below the minimum have larger expected improvements and are therefore more likely to be compositions that could further minimize the current density. Assuming that the current densities at $\boldsymbol{x}$ are normally distributed with mean and standard deviation given by the prediction of the surrogate Gaussian process regressor, the expected improvement can be evaluated as

$$\mathrm{E}[I(\boldsymbol{x})]=(y_{min}+\xi-\mu(\boldsymbol{x}))\,\Phi\left(\frac{y_{min}+\xi-\mu(\boldsymbol{x})}{\sigma(\boldsymbol{x})}\right)+\sigma(\boldsymbol{x})\,\phi\left(\frac{y_{min}+\xi-\mu(\boldsymbol{x})}{\sigma(\boldsymbol{x})}\right), \qquad \text{(S2)}$$

where μ(x) and σ(x) are the mean and standard deviation supplied by the Gaussian process regressor, respectively, and $\phi(t)=(1/\sqrt{2\pi})\exp(-t^2/2)$ and $\Phi(t)=\int_{-\infty}^{t}\phi(t')\,dt'$ are the standard normal probability and cumulative distributions, respectively. The expected improvement was evaluated at the same 1000 compositions as the Gaussian process regressor, and the composition with the maximum acquisition value was further optimized by sampling the expected improvement around the composition in molar fraction steps of 0.005 until a maximum was found that was then selected for sampling by the kinetic model. A $\xi$-value of 0.01 was used throughout. A $\xi$-value of zero was found to potentially discover the locally optimal compositions very quickly. However, discovery of the global optimum was not guaranteed with 150 samples as was the case for $\xi$=0.01.

## Experimental Procedures

### Electrochemical characterization

Binary thin-film composition spreads were analyzed using a high-throughput scanning droplet cell (SDC) coupled with a Jaissle potentiostat/galvanostat. The teflon tip forming the head of the SDC had an opening of 1 mm in diameter, which formed the working electrode in each of the measurement areas (MAs) with a size of $7.35 \cdot 10^{-3}$ $cm^2$, allowing local characterization of the samples. Particular MAs on all of the samples were separated from each other by 2.25 mm, which corresponds to composition changes of ca. 1.5 at.% per element. All electrochemical measurements were conducted in 0.1 M $HClO_4$ electrolyte in a three-electrode system with a Ag|AgCl|3M KCl and a Pt wire as a reference and counter electrode, respectively. Linear sweep voltammetry was performed between 1 V and 200 mV vs. the reversible hydrogen electrode (RHE) with a scan rate of 10 mV $s^{-1}$. All potentials are reported versus the RHE calculated according to the following equation:

$$U_{RHE} = U_{Ag|AgCl|3\,M\,KCl} + 0.210\,V + 0.059\,V \cdot pH\,, \qquad \text{(S3)}$$

where $U_{Ag|AgCl|3M\ KCl}$ is the potential measured vs. the Ag|AgCl|3M KCl reference electrode, 0.210 V is the standard potential of the Ag|AgCl|3M KCl reference electrode at 25 °C. 0.059 V is the result of ln(10)*RT*/*nF*, where *R* is the gas constant, *T* is the temperature (298 K), *F* is the Faraday constant and *n* (=1) is the number of electrons transferred during the reaction.

### Composition analysis

The elemental compositions of all MAs in the MLs were determined using automated energy dispersive X-ray spectroscopy (EDX) at 20 kV acceleration voltage in a scanning electron microscope (SEM, JEOL 5800) using a detector (INCA X-act, Oxford Instruments).

### Surface roughness analysis by AFM

Topographical images of the Ag-Pd, Pd-Ru, Ir-Pt and Ir-(Hi) Pt thin film libraries were measured by atomic force microscopy (AFM, Bruker Dimension Fastscan) using Fastscan mode. For surfaces, whose roughness is characterized by a single length scale, roughness parameters were calculated by the arithmetic mean roughness Ra.

### Phase analysis from XRD

The crystallographic phase analysis was performed using X-ray diffraction (XRD). A Bruker D8 Discover with a Vantec-500 2D-detector in Bragg–Brentano geometry and Cu Kα X-ray source was used. To avoid Si-substrate peaks, measurements were performed in θ–2θ mode with a 2.5° offset on θ. Five frames were taken stepwise at every MA with an increment of θ/2θ 7.5°/15°, starting at 10°/25° and finishing at 40°/85°. In this way an angular 2θ range from approximately 10° to 100° was covered.

### Thin-film fabrication Pd-Ru, Ir-Pt, and Ag-Pd

The Pd-Ru and Ir-Pt libraries were fabricated by a combinatorial magnetron sputtering system (DCA Instruments, Finland) equipped with five cathodes. Two of these five cathodes were positioned at 144° form each other to create composition gradients. High purity (Ir: 99.9%, Pt: 99.99%,) 100 mm diameter single-element targets were used. A confocally-placed 100 mm diameter sapphire wafer (c-plane) was used as a substrate for the Ir-Pt system. It was patterned with small numbered crosses by a photolithographic lift-off process to serve as a reference grid and for making local thickness measurements by stylus profilometry. All of the depositions were carried out without intentional heating. Prior to the deposition, the chamber vacuum was on the order of $10^{-5}$ Pa. During deposition, the pressure was set to 0.667 Pa using Ar (99.9999%) at a flow rate of 60 sccm, and the substrate was kept stationary to obtain continuous compositional gradients. The type of power supply used for each library and sputter powers are listed in Table S1.

The Ag-Pd system was deposited in an alternate vacuum chamber, where cathodes with 38 mm diameter targets (Ag: 99.99%, Pd: 99.95%) are positioned 180° to each other. The substrate used was an approximately 1 cm wide strip cleaved from a 100 mm diameter <100> Si wafer, which was thermally oxidized as a diffusion barrier. The chamber base vacuum was $10^{-4}$ Pa and deposition was done at an Ar pressure of 0.5 Pa. The 100 mm diameter (100) Si substrate with a 500 nm $SiO_2$ barrier layer was stationary at the confocal point of the tilted cathodes so that composition gradients were obtained.

**Table S1.** Sputter parameters for the Pd-Ru, Ir-Pt and Ir-(High)Pt, respectively.

| Libraries | Deposition power (W) | | | |
|---|---|---|---|---|
| | Pd (RF)[a] | Ru (DC)[b] | Ir (DC)[b] | Pt (RF)[a] |
| Pd-Ru | 182 | 44 | - | - |
| Ir-Pt | - | - | 70 | 194 |
| Ir-(Hgh)Pt | - | - | 40 | 239 |

[a] RF: Radio frequency. [b] DC: Direct current.

## Results and Discussion

Number of samples in a grid search of an N-component composition space
The number of combinations of alloy compositions in steps of molar fractions of *s* are given by equation S4.

$$N=\frac{\left(\frac{1}{s}+N_{elems}-1\right)!}{\left(\frac{1}{s}\right)!\left(N_{elems}-1\right)!} \quad \text{(S4)}$$

where $N_{elems}$ is the number of metals in the alloy system. For example, to uniformly span the composition space of a quinary alloy in 5% intervals 10,626 points are needed. Figure S2 shows the number of combinations needed to span the composition space for various steps of molar fractions. It is observed that as the number of elements increases, the exploration of the composition space becomes increasingly infeasible as the number of samples needed increase combinatorially.

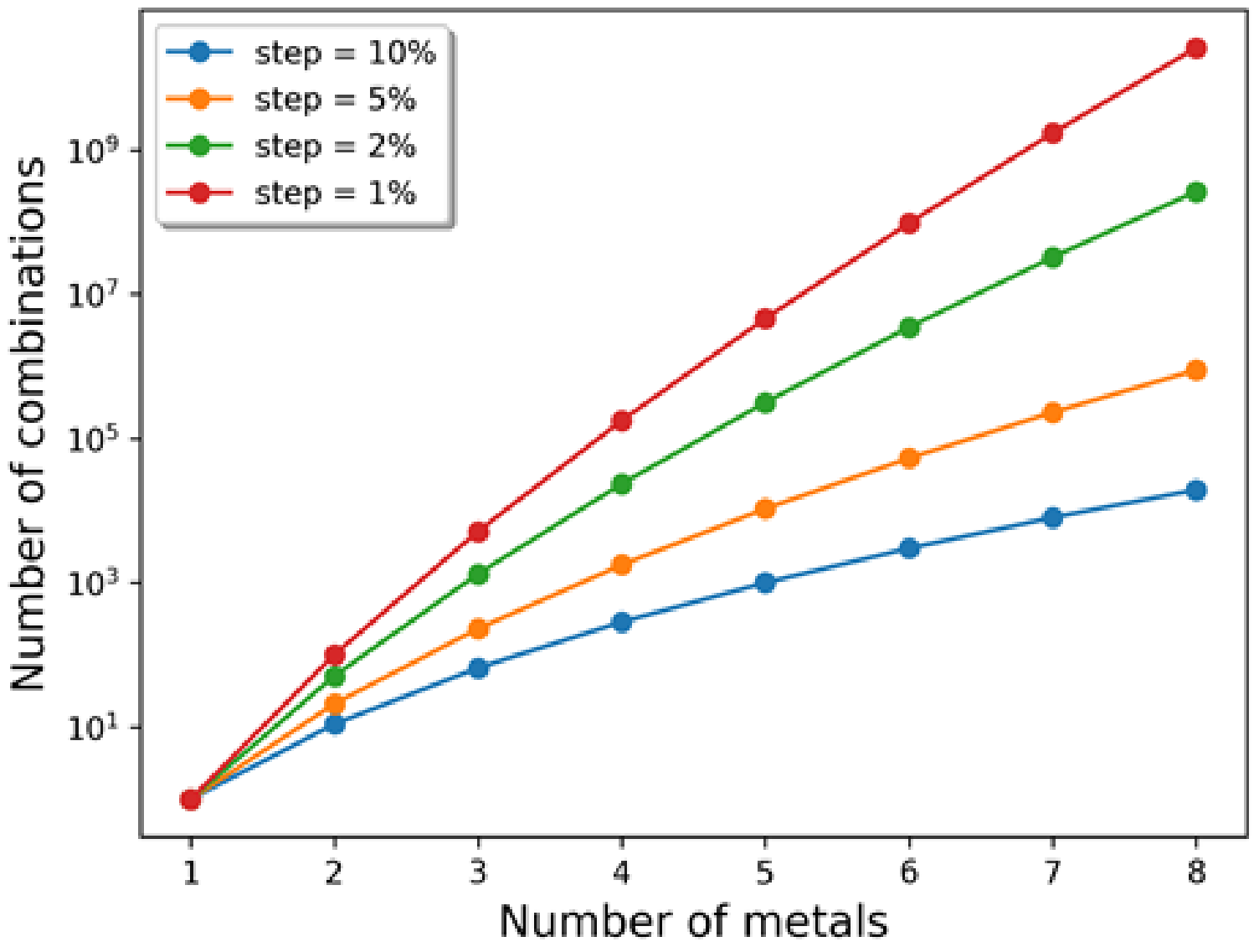

**Figure S2.** Number of samples needed to span the composition space. Shown for molar fraction step sizes of 1, 2, 5 and 10% as a function of the number of metals in the alloy.

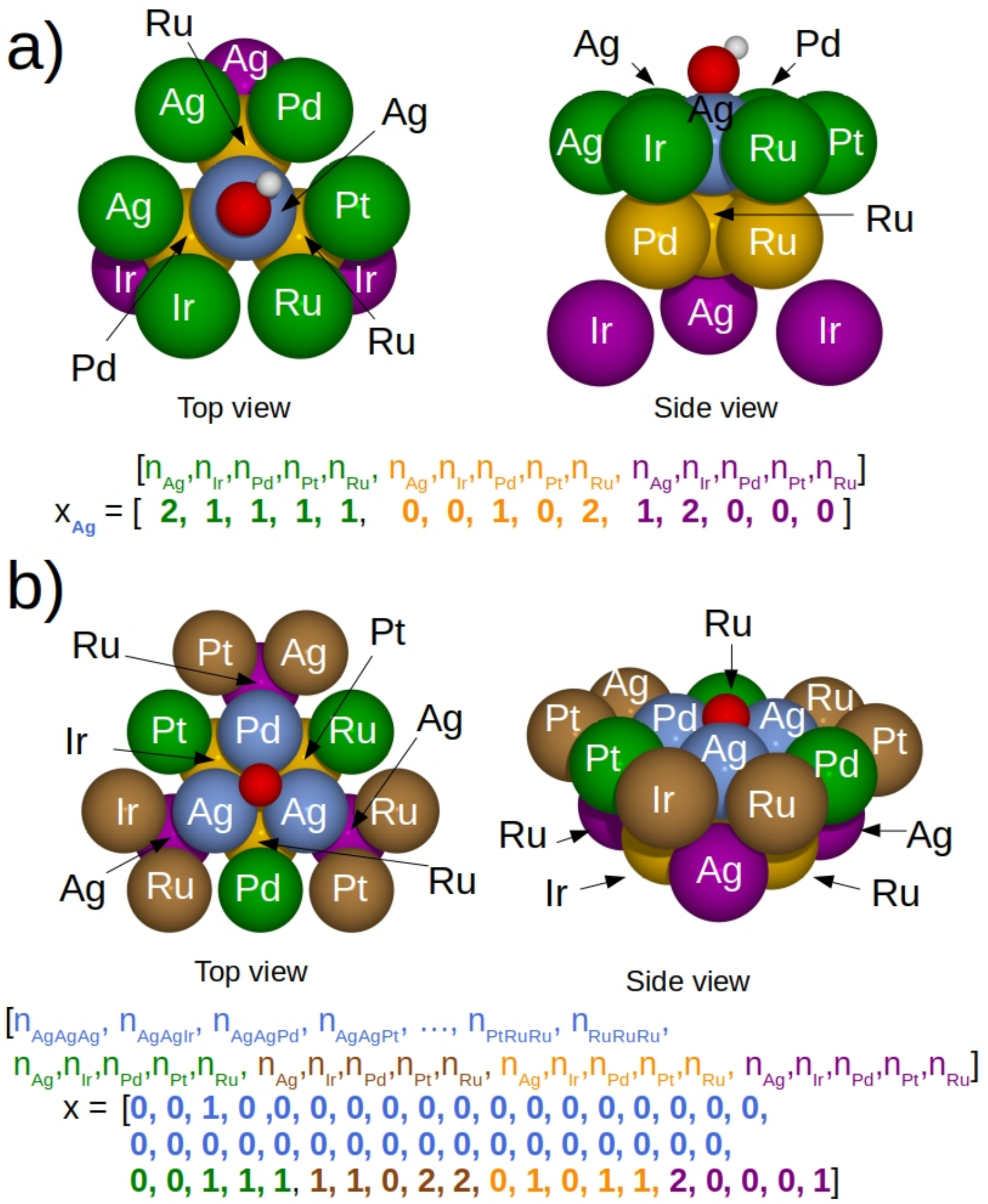

**Figure S3.** Examples of encoding the features of a structure. Shown in **a)** for an *OH on-top site with a set of features for each possible on-top site element (exemplified for an Ag on-top site) with a total of 15 parameters, and in **b)** for an O* fcc hollow site with one-hot encoding of the adsorption site ensemble (here exemplified for the AgAgPd ensemble) with a total of 55 parameters. The color of the text matches the corresponding colors of the atoms in the structure.

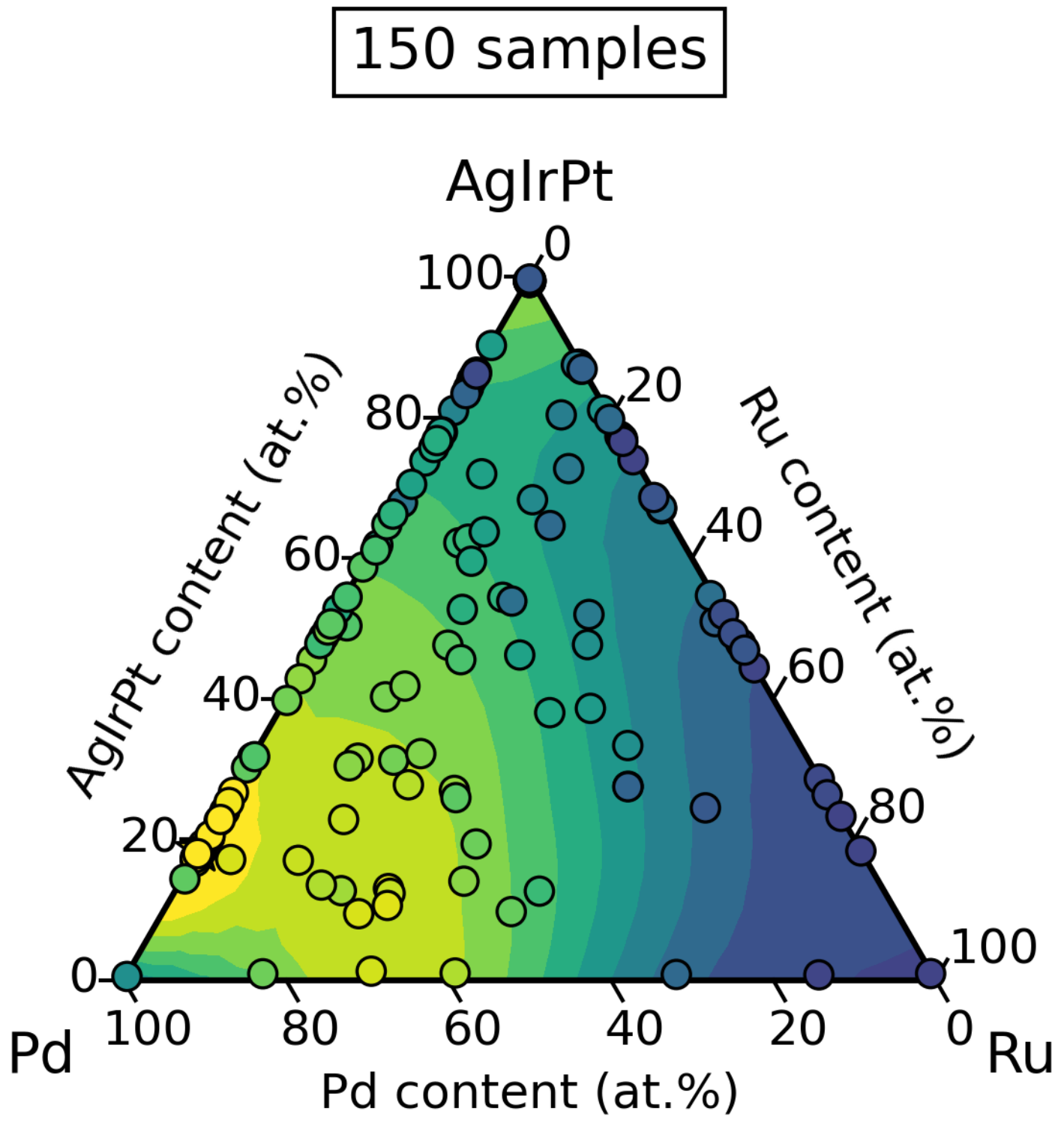

**Figure S4.** Pseudo-ternary plot of the Ag-Ir-Pd-Pt-Ru modeled current densities after sampling of 150 samples with Bayesian optimization. The Ag, Ir, and Pt concentrations have been grouped into one to highlight the plateau of similar current densities when the binary $Pd_{65}Ru_{35}$ is mixed with other elements in trace amounts. Yellow colors correspond to regions with high absolute values of the current density and blue colors to correspondingly low values. The projection of the current density from the quinary to the pseudo-ternary composition space was accomplished by showing the maximal absolute value of the current density possible for compositions that would otherwise be overlapping.

**Table S2.** Linear parameters used for on-top *OH adsorption energy prediction on the Ag-Ir-Pd-Pt-Ru system. The parameters will give the electronic energy in eV relative to *OH on Pt(111). The order of the parameters, after the intercept, follows the features given in Figure S3a. In the labels of the parameters the number refers to the layer, and the letter to the proximity to the adsorption site, e.g. "3a Pd" refers to the influence of Pd in the nearest atoms in the 3rd layer below the surface. The intercept has been chosen to yield the prediction for the pure element, since this value is obtained by setting the respective parameters for that element in each zone to zero.

| **Ag-Ir-Pd-Pt-Ru** (eV relative to *OH@Pt(111)) | | | | | |
|---|---|---|---|---|---|
| | **@Ag** | **@Ir** | **@Pd** | **@Pt** | **@Ru** |
| **Intercept** | 0.262515 | -0.363696 | 0.075474 | -0.023367 | -0.604279 |
| **1b Ag** | 0 | -0.074370 | -0.025064 | -0.065560 | -0.026962 |
| **1b Ir** | 0.124874 | 0 | 0.044604 | 0.014894 | 0.025734 |
| **1b Pd** | 0.055553 | -0.057413 | 0 | -0.042658 | -0.021608 |
| **1b Pt** | 0.106736 | -0.016379 | 0.040827 | 0 | 0.021586 |
| **1b Ru** | 0.083629 | -0.019943 | 0.007435 | -0.016870 | 0 |
| **2a Ag** | 0 | 0.020613 | 0.022701 | 0.076312 | -0.097415 |
| **2a Ir** | -0.103170 | 0 | -0.050573 | -0.022161 | -0.022948 |
| **2a Pd** | -0.025039 | 0.008149 | 0 | 0.046161 | -0.070458 |
| **2a Pt** | -0.057300 | -0.012203 | -0.032788 | 0 | -0.059442 |
| **2a Ru** | -0.121723 | 0.031539 | -0.051361 | -0.014583 | 0 |
| **3a Ag** | 0 | -0.036480 | -0.015738 | -0.022794 | -0.025380 |
| **3a Ir** | 0.019881 | 0 | 0.035545 | 0.032103 | 0.013630 |
| **3a Pd** | -0.001190 | -0.014928 | 0 | -0.006961 | -0.000195 |
| **3a Pt** | 0.005926 | -0.012114 | 0.000055 | 0 | 0.008633 |
| **3a Ru** | 0.023315 | -0.002794 | 0.033945 | 0.039372 | 0 |

**Table S3.** Linear parameters used for fcc hollow O* adsorption energy prediction on the Ag-Ir-Pd-Pt-Ru system. The parameters will give the electronic energy in eV relative to O* on Pt(111). The order of the parameters follows the features given in Figure S3b. In the labels of the parameters the number refers to the surface layer, and the letter to the proximity to the adsorption site, e.g. “2b Pd” refers to the influence of Pd in the next nearest atoms in the subsurface layer.

**Ag-Ir-Pd-Pt-Ru** (eV relative to O*@Pt(111))

| | | | | | | | |
|---|---|---|---|---|---|---|---|
| **AgAgAg** | 0.823301 | **AgAgIr** | -0.470179 | **AgAgPd** | 0.612573 | **AgAgPt** | 0.374752 |
| **AgAgRu** | -0.869420 | **AgIrIr** | -0.603051 | **AgIrPd** | -0.386478 | **AgIrPt** | -0.296955 |
| **AgAgRu** | -0.971134 | **AgPdPd** | 0.417371 | **AgPdPt** | 0.226634 | **AgPdRu** | -0.897300 |
| **AgPtPt** | 0.231177 | **AgPtRu** | -0.677207 | **AgRuRu** | -1.168980 | **IrIrIr** | -0.884791 |
| **IrIrPd** | -0.791251 | **IrIrPt** | -0.645038 | **IrIrRu** | -1.154468 | **IrPdPd** | -0.586752 |
| **IrPdPt** | -0.435780 | **IrPdRu** | -1.101158 | **IrPtPt** | -0.331185 | **IrPtRu** | -0.912780 |
| **IrRuRu** | -1.396306 | **PdPdPd** | 0.142790 | **PdPdPt** | 0.030062 | **PdPdRu** | -0.915004 |
| **PdPtPt** | 0.073705 | **PdPtRu** | -0.771205 | **PdRuRu** | -1.356419 | **PtPtPt** | 0.092753 |
| **PtPtRu** | -0.617444 | **PtRuRu** | -1.178394 | **RuRuRu** | -1.654559 | - | - |
| **1b Ag** | -0.081909 | **1c Ag** | 0.028809 | **2a Ag** | -0.040527 | **2B Ag** | 0.005127 |
| **1b Ir** | 0.062988 | **1c Ir** | -0.020630 | **2a Ir** | 0.030518 | **2b Ir** | -0.004639 |
| **1b Pd** | -0.047180 | **1c Pd** | 0.014160 | **2a Pd** | -0.022949 | **2b Pd** | -0.013672 |
| **1b Pt** | 0.025269 | **1c Pt** | 0.002563 | **2a Pt** | -0.011230 | **2b Pt** | -0.027344 |
| **1b Ru** | 0.040649 | **1c Ru** | -0.025635 | **2a Ru** | 0.042969 | **2b Ru** | 0.040405 |

**Table S4.** Linear parameters used for on-top *OH adsorption energy prediction on the Ir-Pd-Pt-Rh-Ru system. The parameters will give the electronic energy in eV relative to *OH on Pt(111). The order of the parameters, after the intercept, follows the features given in Figure S3a: In the labels of the parameters the number refers to the layer, and the letter to the proximity to the adsorption site, e.g. “3a Pd” refers to the influence of Pd in the nearest atoms in the 3rd layer below the surface. The intercept has been chosen to yield the prediction for the pure element, since this value is obtained by setting the respective parameters for that element in each zone to zero.

| **Ir-Pd-Pt-Rh-Ru** (eV relative to *OH@Pt(111)) | | | | | |
|---|---|---|---|---|---|
| | **@Ir** | **@Pd** | **@Pt** | **@Rh** | **@Ru** |
| **Intercept** | -0.324264 | 0.044878 | -0.008922 | -0.323476 | -0.639564 |
| **1b Ir** | 0 | 0.045847 | 0.011095 | 0.037634 | 0.040998 |
| **1b Pd** | -0.067114 | 0 | -0.041923 | -0.012061 | -0.024511 |
| **1b Pt** | -0.022217 | 0.041869 | 0 | 0.028185 | 0.018334 |
| **1b Rh** | -0.048245 | 0.006241 | -0.030170 | 0 | -0.002666 |
| **1b Ru** | -0.026561 | 0.001899 | -0.020977 | 0.002107 | 0 |
| **2a Ir** | 0 | -0.034497 | -0.013930 | -0.007092 | -0.018194 |
| **2a Pd** | 0.017262 | 0 | 0.030018 | -0.006571 | -0.047992 |
| **2a Pt** | -0.006588 | -0.027407 | 0 | -0.014891 | -0.037786 |
| **2a Rh** | 0.009353 | -0.020069 | 0.021633 | 0 | -0.022024 |
| **2a Ru** | 0.021847 | -0.030915 | 0.002223 | 0.005128 | 0 |
| **3a Ir** | 0 | 0.034227 | 0.029385 | 0.006703 | 0.004741 |
| **3a Pd** | -0.026733 | 0 | -0.016174 | -0.021482 | -0.015521 |
| **3a Pt** | -0.016911 | 0.012355 | 0 | -0.006625 | -0.000825 |
| **3a Rh** | -0.005591 | 0.015988 | 0.015289 | 0 | -0.003526 |
| **3a Ru** | 0.006793 | 0.035460 | 0.039082 | 0.007343 | 0 |

**Table S5.** Linear parameters used for fcc hollow O* adsorption energy prediction on the Ir-Pd-Pt-Rh-Ru system. The parameters will give the electronic energy in eV relative to O* on Pt(111). The order of the parameters follows the features given in Figure S3b. In the labels of the parameters the number refers to the surface layer, and the letter to the proximity to the adsorption site, e.g. "2b Pd" refers to the influence of Pd in the next nearest atoms in the subsurface layer.

| **Ir-Pd-Pt-Rh-Ru** (eV relative to O*@Pt(111)) | | | | | | | |
|---|---|---|---|---|---|---|---|
| **IrIrIr** | -0.802671 | **IrIrPd** | -0.698724 | **IrIrPt** | -0.549723 | **IrIrRh** | -0.867686 |
| **IrIrRu** | -1.039895 | **IrPdPd** | -0.490209 | **IrPdPt** | -0.372960 | **IrPdRh** | -0.717682 |
| **IrIrRu** | -1.023565 | **IrPtPt** | -0.244375 | **IrPtRh** | -0.585337 | **IrPtRu** | -0.815201 |
| **IrRhRh** | -0.894818 | **IrRhRu** | -1.128900 | **IrRuRu** | -1.293938 | **PdPdPd** | 0.154385 |
| **PdPdPt** | 0.096998 | **PdPdRh** | -0.273295 | **PdPdRu** | -0.866591 | **PdPtPt** | 0.062937 |
| **PdPtRh** | -0.252892 | **PdPtRu** | -0.716471 | **PdRhRh** | -0.602280 | **PdRhRu** | -1.054314 |
| **PdRuRu** | -1.262522 | **PtPtPt** | 0.183877 | **PtPtRh** | -0.153542 | **PtPtRu** | -0.560802 |
| **PtRhRh** | -0.506731 | **PtRhRu** | -0.874115 | **PtRuRu** | -1.077222 | **RhRhRh** | -0.856961 |
| **RhRhRu** | -1.144941 | **RhRuRu** | -1.392573 | **RuRuRu** | -1.563099 | - | - |
| **1b Ir** | 0.045966 | **1c Ir** | -0.020285 | **2a Ir** | 0.026245 | **2B Ir** | 0.002915 |
| **1b Pd** | -0.061963 | **1c Pd** | 0.029724 | **2a Pd** | -0.040319 | **2b Pd** | -0.028029 |
| **1b Pt** | 0.006466 | **1c Pt** | 0.012666 | **2a Pt** | -0.022935 | **2b Pt** | -0.039050 |
| **1b Rh** | -0.011702 | **1c Rh** | 0.000303 | **2a Rh** | 0.007976 | **2b Rh** | 0.003210 |
| **1b Ru** | 0.021233 | **1c Ru** | -0.022409 | **2a Ru** | 0.029033 | **2b Ru** | 0.060954 |

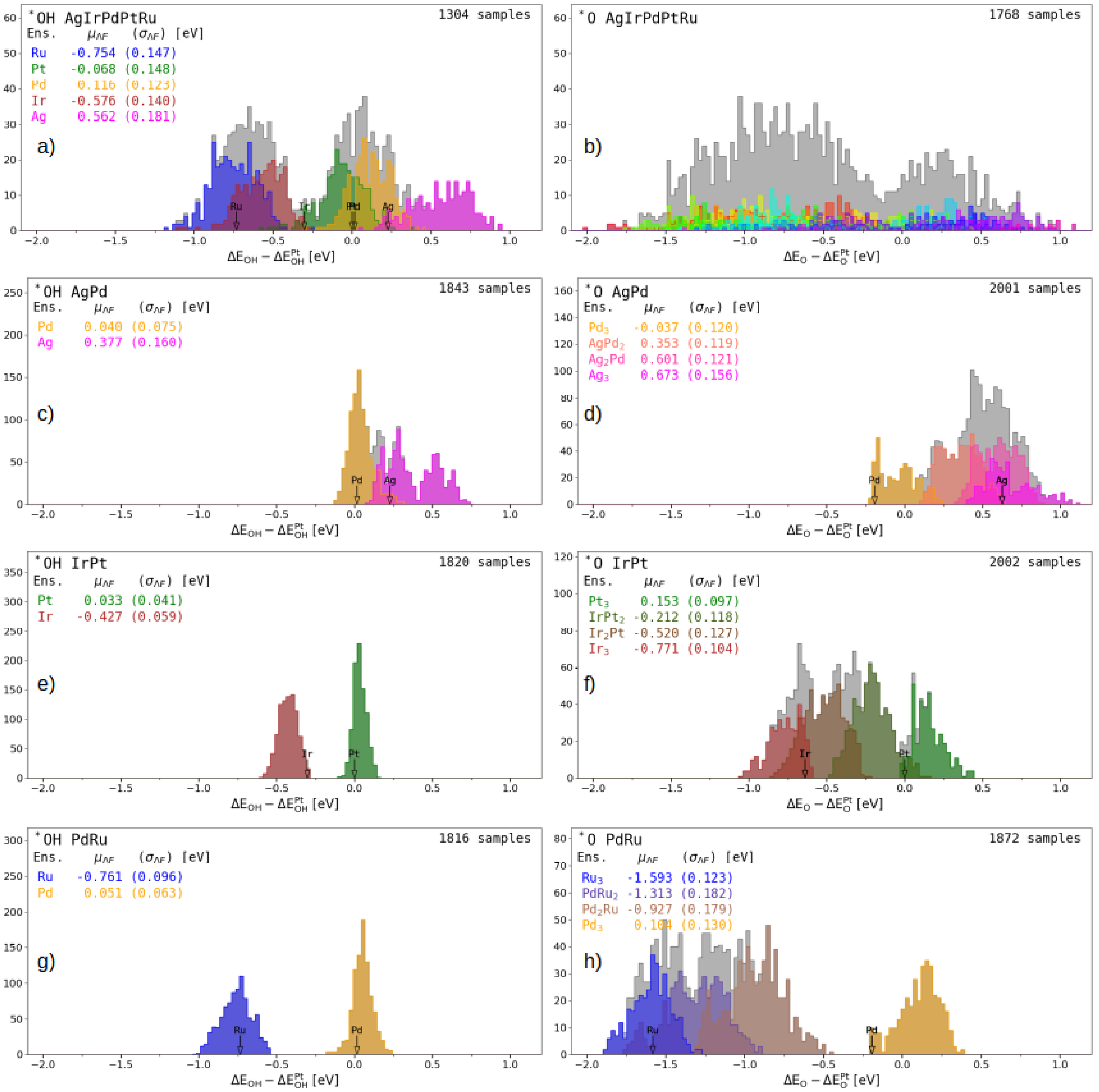

**Figure S5.** Histograms of DFT calculated adsorption energies of *OH and *O on the quinary alloy and the three binary alloys. Adsorption ensembles are distinguished by different colors with the mean adsorption energy ($\mu_{\Delta E}$) and standard deviation ($\sigma_{\Delta E}$). Adsorption energies of the pure metal fcc(111) surfaces are marked with black arrows.

**Table S6.** Mean absolute errors (MAEs) in units of eV of several regression algorithms predicting adsorption energies of *OH and O* on the quinary and binary alloys using truncated site features as displayed in Figure S3. 20% of the samples were selected for testing with the remaining samples used to train the regression model. The standard deviations on the last digit(s) of the MAEs are displayed in parentheses.

*OH adsorption energies - Truncated adsorption site features

| Regressor type | Ag-Pd (1843 samples) | Ir-Pt (1820 samples) | Pd-Ru (1816 samples) | Ag-Ir-Pd-Pt-Ru (1304 samples) |
|---|---|---|---|---|
| Dummy (mean) | 0.09(7) | 0.05(4) | 0.06(5) | 0.12(9) |
| Linear regr. | 0.04(3) | 0.02(2) | 0.05(4) | 0.06(5) |
| Ridge regr. | 0.04(3) | 0.02(2) | 0.05(4) | 0.06(5) |
| Gradient Boosting | 0.03(2) | 0.018(14) | 0.03(3) | 0.07(6) |
| Random Forest | 0.03(2) | 0.018(15) | 0.04(3) | 0.07(6) |

*O adsorption energies - Truncated adsorption site features

| Regressor type | Ag-Pd (2001 samples) | Ir-Pt (2002 samples) | Pd-Ru (1872 samples) | Ag-Ir-Pd-Pt-Ru (1768 samples) |
|---|---|---|---|---|
| Dummy (mean) | 0.11(8) | 0.10(8) | 0.15(10) | 0.5(4) |
| Linear regr. | 0.05(4) | 0.04(3) | 0.08(6) | 0.09(7) |
| Ridge regr. | 0.05(4) | 0.04(3) | 0.08(5) | 0.11(8) |
| Gradient Boosting | 0.04(3) | 0.04(3) | 0.05(4) | 0.11(8) |
| Random Forest | 0.03(3) | 0.04(3) | 0.05(4) | 0.10(8) |

**Table S7.** Mean absolute errors (MAEs) in units of eV of several regression algorithms predicting adsorption energies of *OH and O* on the quinary and binary alloys using extended site features (up to fourth-nearest neighboring atoms for all layers). 20% of the samples were selected for testing with the remaining samples used to train the regression model. The standard deviations on the last digit(s) of the MAEs are displayed in parentheses.

*OH adsorption energies - Extended adsorption site features

| Regressor type | Ag-Pd (1843 samples) | Ir-Pt (1820 samples) | Pd-Ru (1816 samples) | Ag-Ir-Pd-Pt-Ru (1304 samples) |
|---|---|---|---|---|
| Dummy (mean) | 0.09(7) | 0.05(4) | 0.06(5) | 0.12(9) |
| Linear regr. | 0.04(3) | 0.020(17) | 0.05(4) | 0.08(7) |
| Ridge regr. | 0.03(3) | 0.020(19) | 0.04(3) | 0.06(5) |
| Gradient Boosting | 0.017(16) | 0.012(9) | 0.03(2) | 0.07(5) |
| Random Forest | 0.015(13) | 0.011(10) | 0.03(2) | 0.07(6) |

*O adsorption energies - Extended adsorption site features

| Regressor type | AgPd (2001 samples) | IrPt (2002 samples) | PdRu (1872 samples) | AgIrPdPtRu (1768 samples) |
|---|---|---|---|---|
| Dummy (mean) | 0.11(8) | 0.10(8) | 0.15(10) | 0.5(4) |
| Linear regr. | 0.04(4) | 0.02(2) | 0.06(4) | 0.09(7) |
| Ridge regr. | 0.05(3) | 0.026(19) | 0.06(4) | 0.10(8) |
| Gradient Boosting | 0.024(19) | 0.021(17) | 0.04(3) | 0.08(6) |
| Random Forest | 0.020(18) | 0.024(19) | 0.03(3) | 0.09(8) |

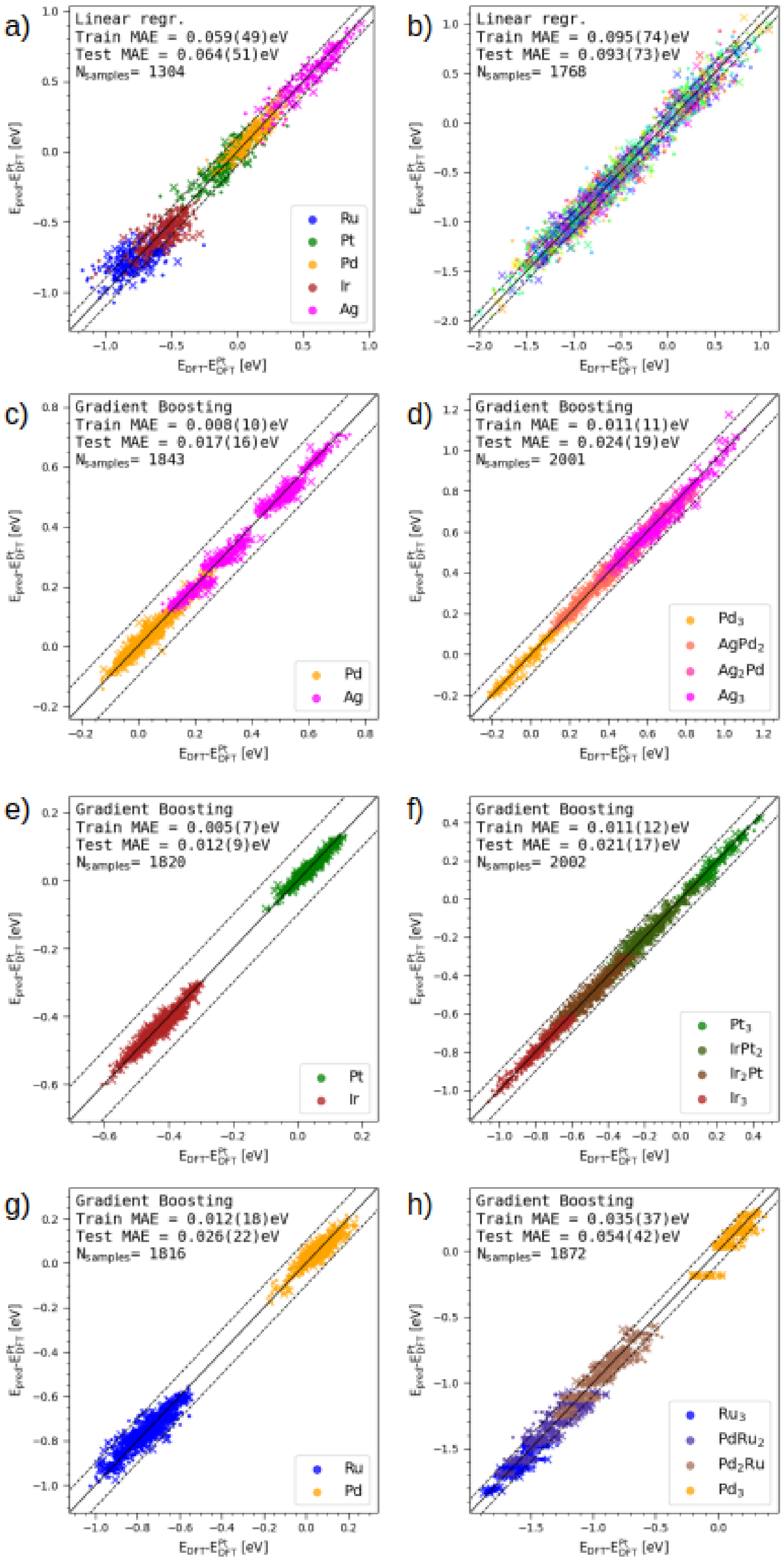

**Figure S6.** Predicted adsorption energy plotted against DFT calculated adsorption energy for the Ag-Ir-Pd-Pt-Ru linear model (**a**,**b**), Ag-Pd (**c**,**d**), Ir-Pt (**e**,**f**), and Pd-Ru (**g**,**h**) gradient boosted models for on-top *OH (**a**,**c**,**e**,**g**) and fcc hollow O* (**b**,**d**,**f**,**h**) adsorption on fcc(111) surfaces. The colors indicate the identity of the adsorption site as shown in the legend. Mean absolute errors with standard deviations are displayed for both training and test set. 20% of the samples were selected for testing (crosses) with the remaining samples used to train the model (circles). For the quinary alloy the results of the linear regression model trained on the truncated site features are displayed, while for the binary alloys the results of the gradient boosted model trained on the extended site features are displayed.

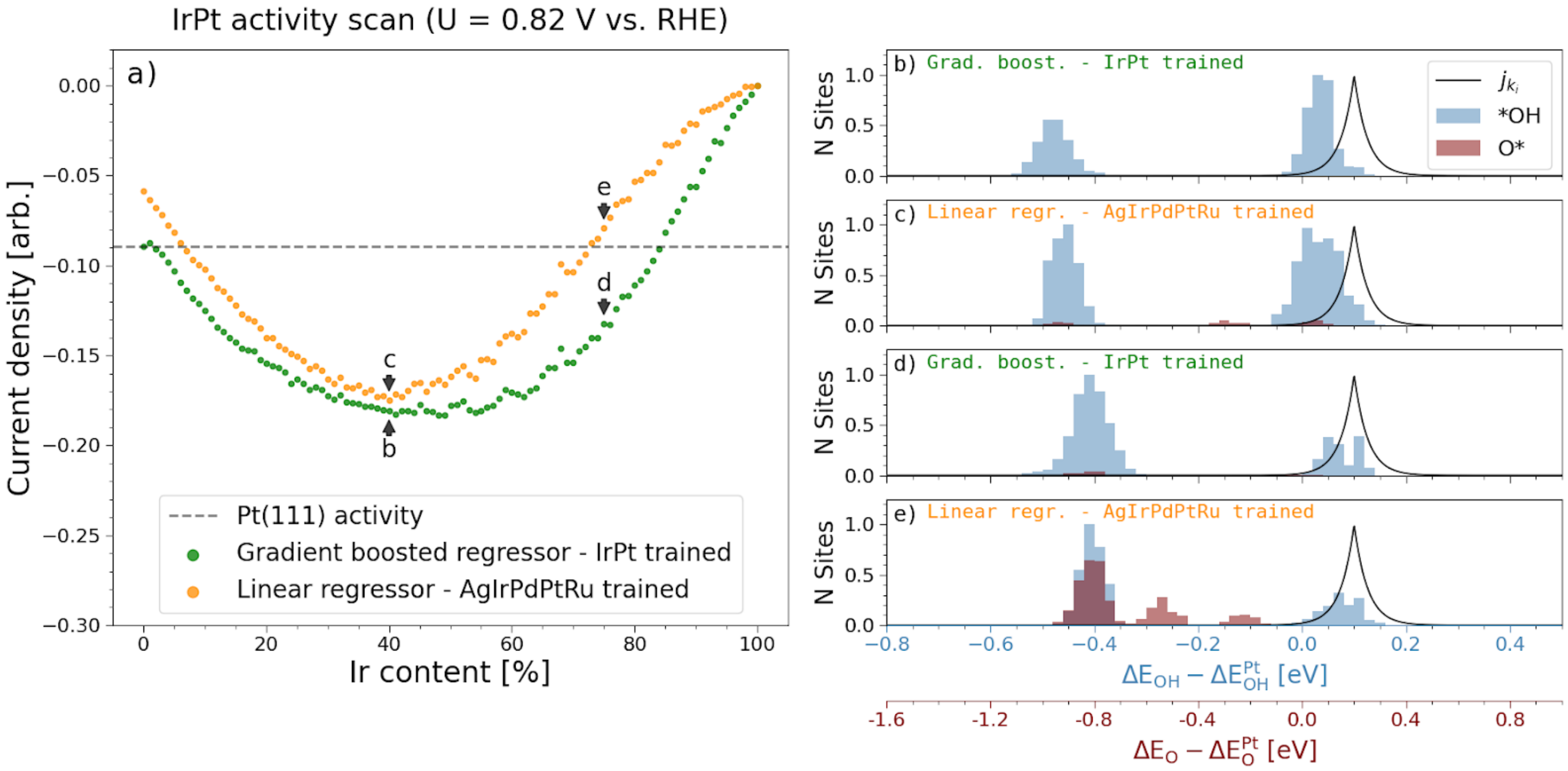

**Figure S7.** Simulated current densities of the Ir-Pt binary system shown as a scan from pure Pt to pure Ir with 1 at.% increments. A linear regression model trained on the DFT calculated samples of the quinary alloy is used alongside a gradient boosted model trained on DFT calculated samples of IrPt to predict the adsorption energies of the simulated surface. These predictions serve as input for Equations 1-3 which yield the resulting current densities. Net adsorption energy distributions (including intersite blocking) are displayed for select composition on the right. The histograms are normalized against the largest bincount and a scaled visualization of the $j_{k,i}$ term is inserted.

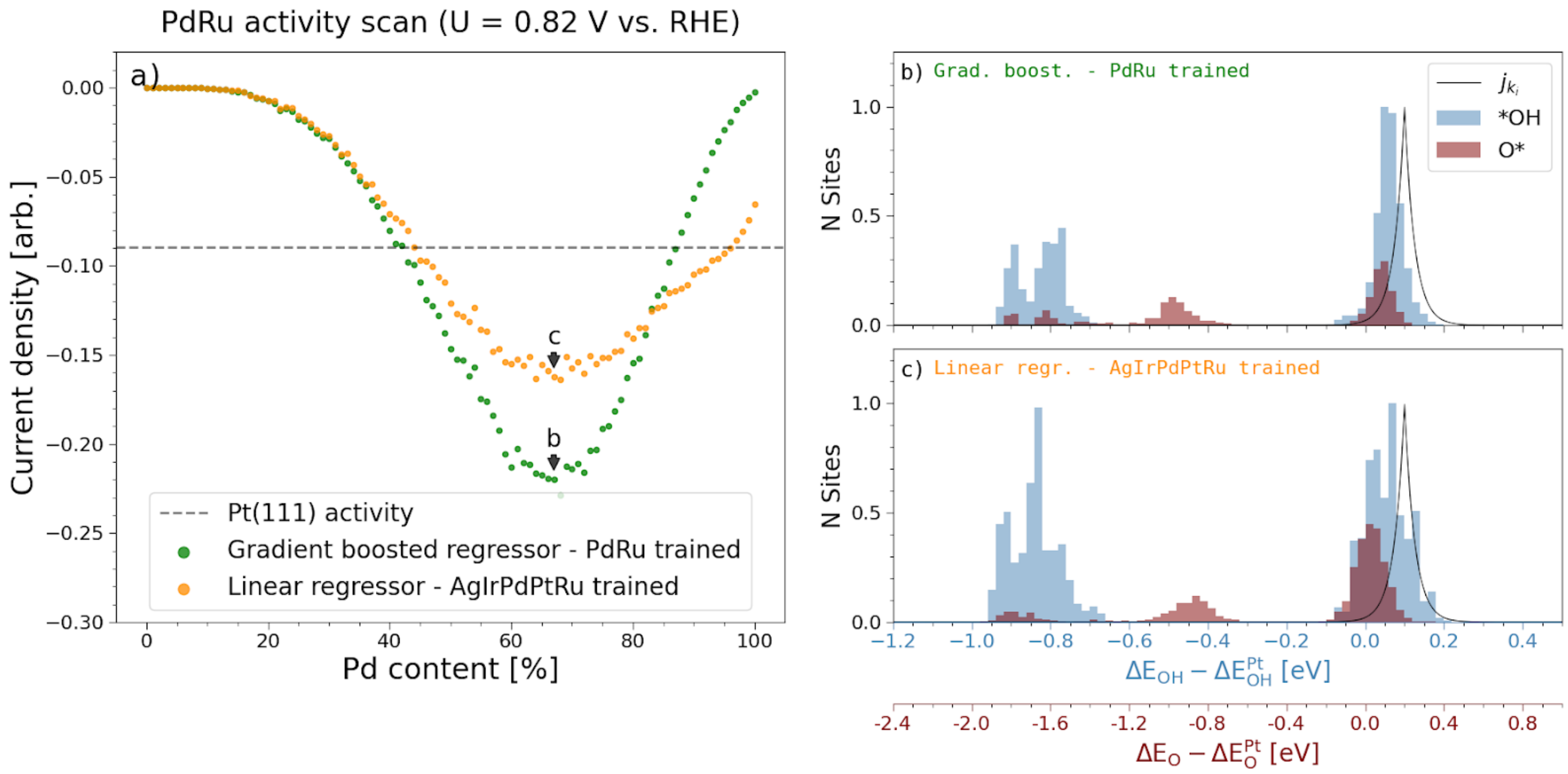

**Figure S8.** Simulated current densities of the Pd-Ru binary system shown as a scan from pure Ru to pure Pd with 1 at.% increments. A linear regression model trained on the DFT calculated samples of the quinary alloy is used alongside a gradient boosted model trained on DFT calculated samples of PdRu to predict the adsorption energies of the simulated surface. These predictions serve as input for Equations 1-3 which yield the resulting current densities. Net adsorption energy distributions (including intersite blocking) are displayed for select composition on the right. The histograms are normalized against the largest bincount and a scaled visualization of the $j_{k,i}$ term is inserted.

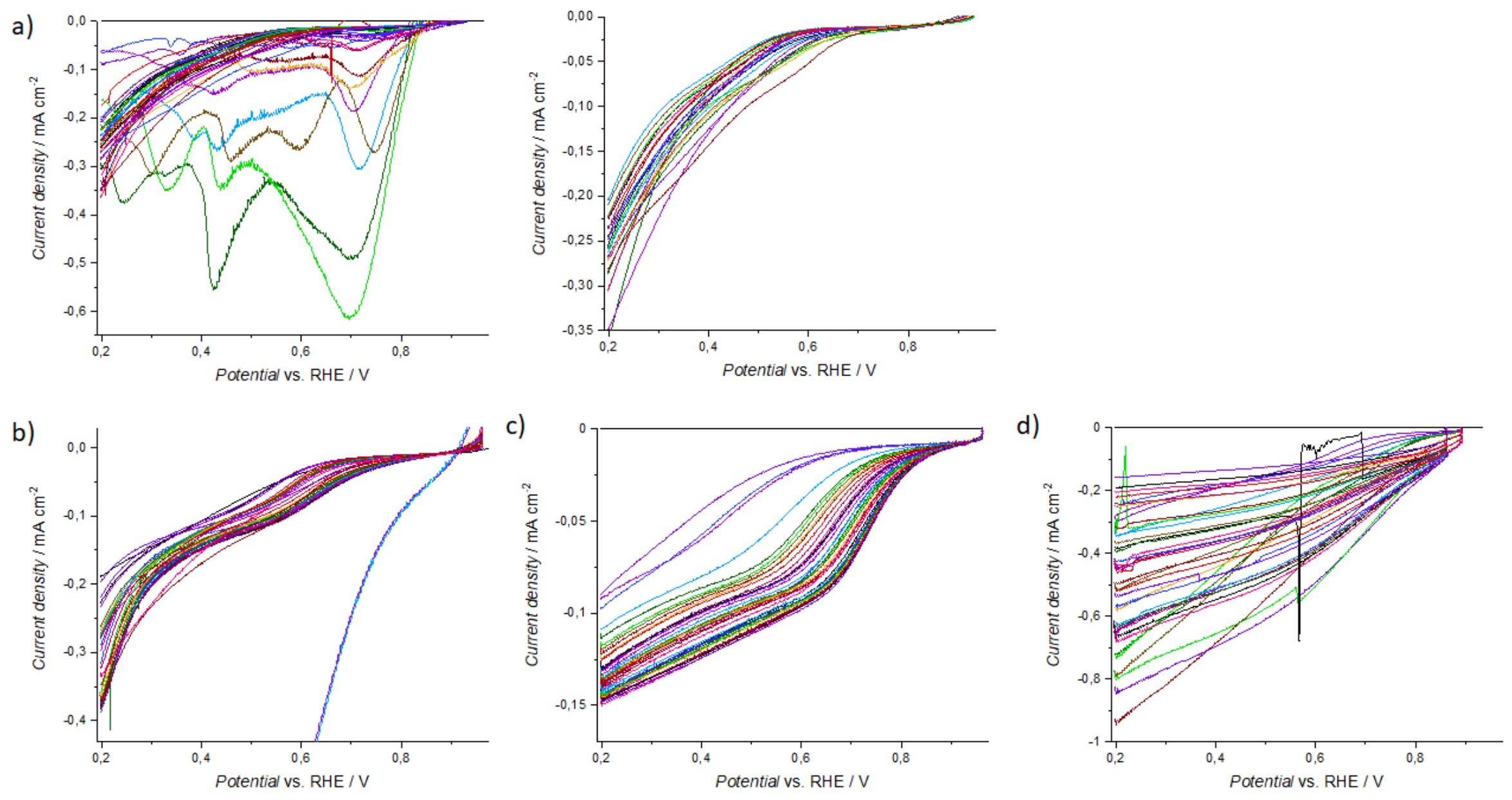

**Figure S9.** All automatically measured LSV curves. **a)** Ag-Pd (right side: LSV plots from the low-Ag part of the sample, without visible film corrosion), **b)** Pd-Ru, and **c)** and **d)** Ir-Pt binaries.

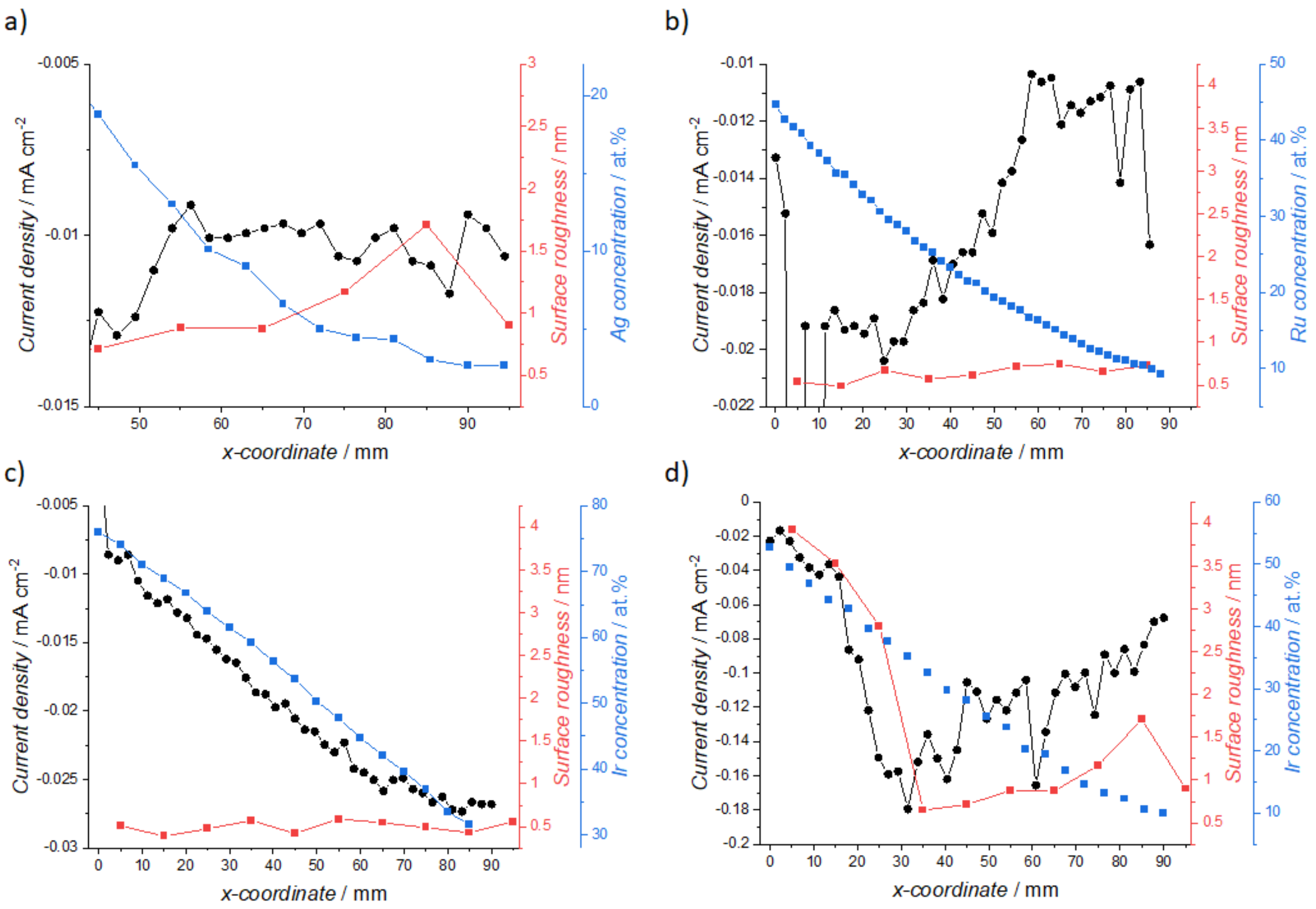

**Figure S10.** Comparisons of measured ORR current densities (black curve) with sample composition (blue curve) and surface roughness (red curve) for synthesized thin-films of **a)** Ag-Pd, **b)** Pd-Ru, **c)** Ir-Pt, and **d)** Ir-(High)Pt.

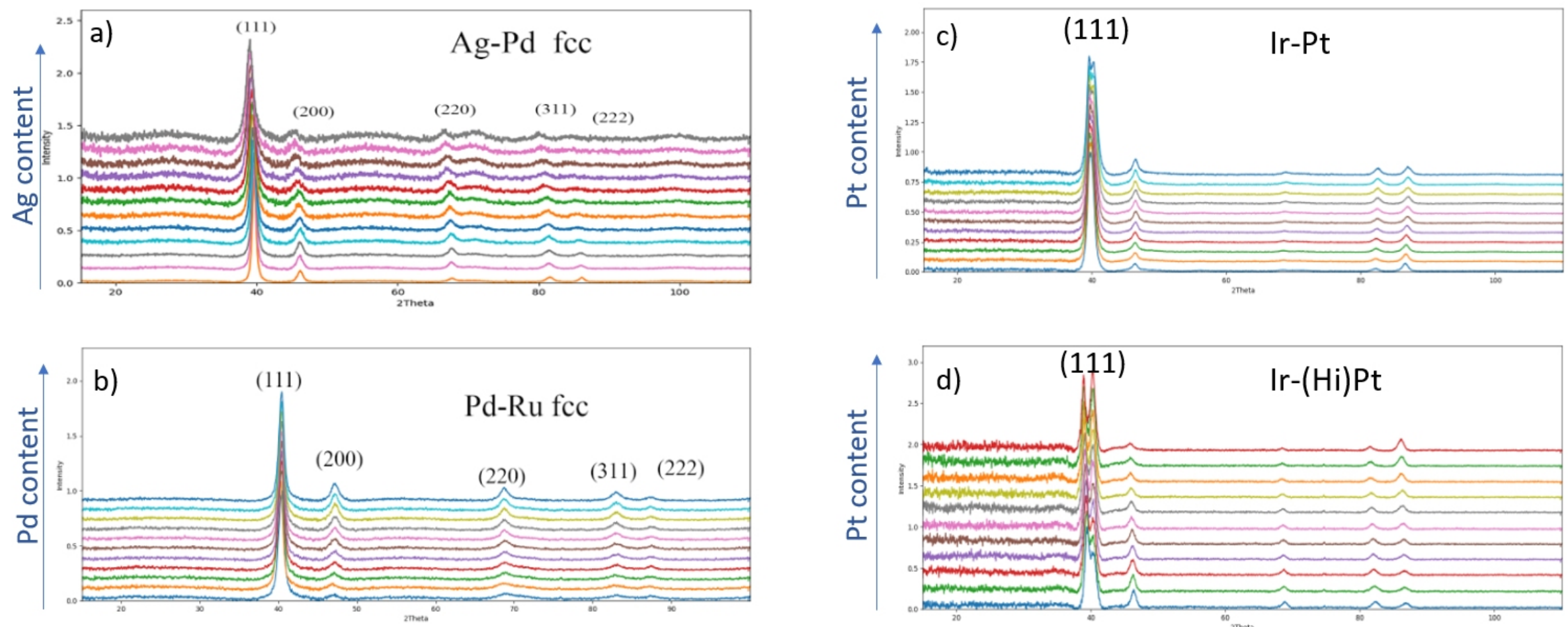

**Figure S11.** XRD profiles for **a)** Ag-Pd, **b)** Pd-Ru **c)** Ir-Pt, and **d)** Ir-(High)Pt binaries. A single fcc phase is observed for both the Ag-Pd and Pd-Ru systems, regardless of composition, while a dual phase is found for the Ir-Pt system.